\pdfoutput=1

\documentclass[11pt]{article}

\usepackage[preprint]{coling}

\usepackage{times}
\usepackage{latexsym}
\usepackage{float}
\usepackage{graphicx}
\usepackage{amsmath}
\usepackage{multirow}
\usepackage{booktabs}
\usepackage{rotating}
\usepackage{subcaption}
\usepackage{tcolorbox}
\usepackage{threeparttable}
\usepackage{pythonhighlight}
\usepackage{listings}
\usepackage{algorithm}
\usepackage{algorithmic}
\usepackage{tabularx}
\usepackage{makecell}
\usepackage{enumitem}
\usepackage{adjustbox}

\usepackage[T1]{fontenc}

\usepackage[utf8]{inputenc}

\usepackage{microtype}

\usepackage{inconsolata}

\usepackage{graphicx}
\usepackage{longtable}
\usepackage{colortbl}

\newcommand{\CJemoji}{\includegraphics[height=1.7\fontcharht\font`\B]{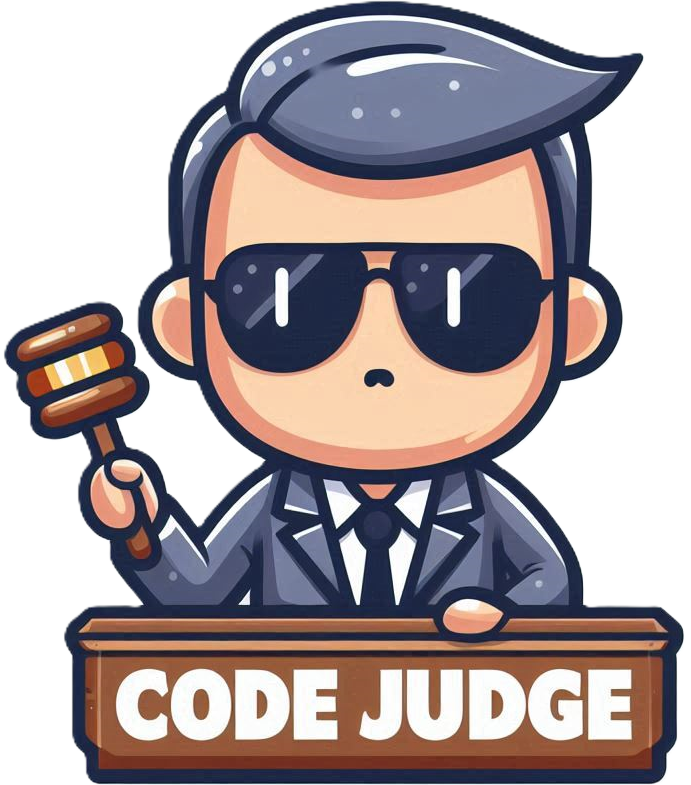}}
\newcommand{\Metaemoji}{\includegraphics[height=0.8\fontcharht\font`\B]{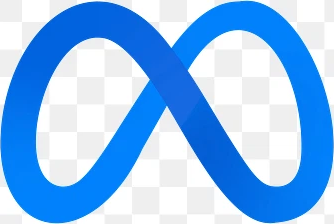}}
\newcommand{\Googleemoji}{\includegraphics[height=1.2\fontcharht\font`\B]{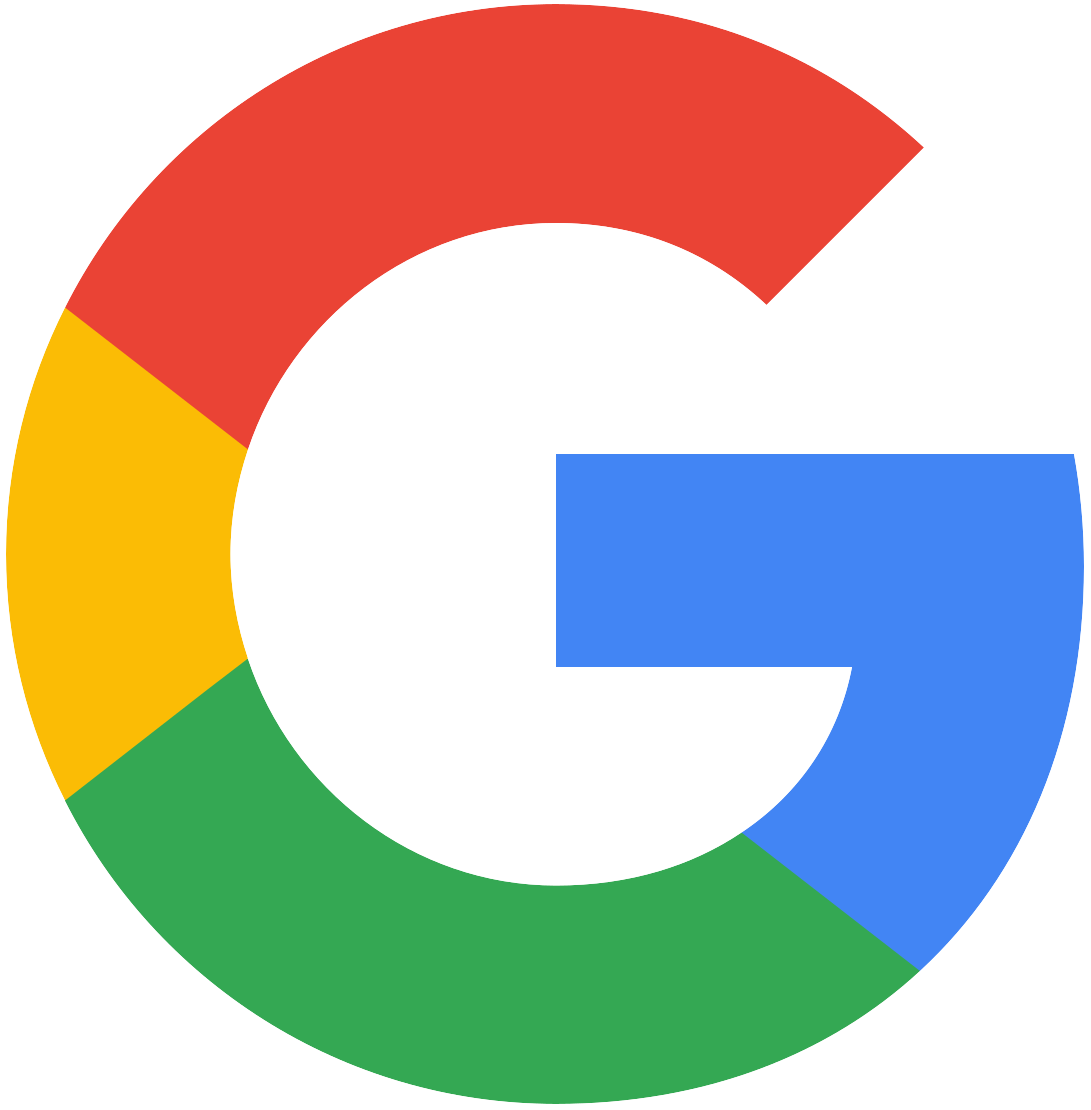}}
\newcommand{\Mistralemoji}{\includegraphics[height=1.2\fontcharht\font`\B]{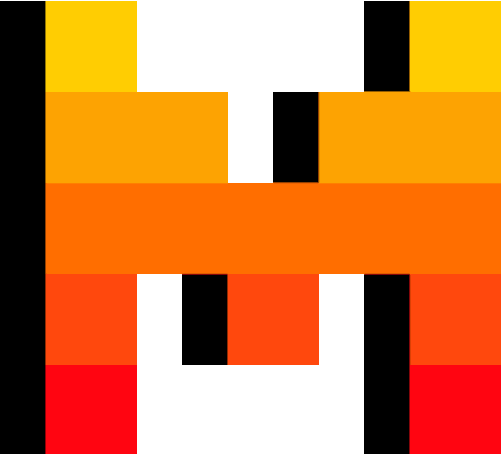}}
\newcommand{\Openaiemoji}{\includegraphics[height=1.2\fontcharht\font`\B]{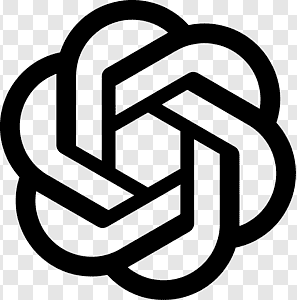}}
\newcommand{\Anthropicemoji}{\includegraphics[height=0.8\fontcharht\font`\B]{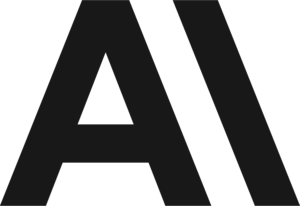}}
\newcommand{\Qwenemoji}{\includegraphics[height=1.2\fontcharht\font`\B]{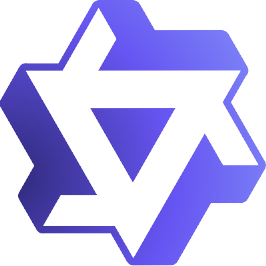}}
\newcommand{\Deepseekemoji}{\includegraphics[height=\fontcharht\font`\B]{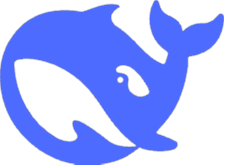}}

\title{\CJemoji{} CodeJudge-Eval:  Can Large Language Models be Good Judges in\\Code Understanding?}

\author{%
  $^{\diamondsuit}$Yuwei Zhao\thanks{Equal Contribution. Ziyang Luo is the project lead.}~,
  $^{\spadesuit}$Ziyang Luo\footnotemark[1]~,
  $^{\heartsuit}$Yuchen Tian~,
  $^{\spadesuit}$Hongzhan Lin~\\[5pt]
  $^{\clubsuit}$Weixiang Yan~,
  $^{\diamondsuit}$Annan Li~,
  $^{\spadesuit}$Jing Ma\thanks{Corresponding Author.}\\[5pt]
  $^\spadesuit$Hong Kong Baptist University, $^\diamondsuit$Beihang University\\[3pt]
  $^\heartsuit$University of Tokyo,
  $^\clubsuit$Vaneval.AI\\[3pt]
  \texttt{\{yuweizhao,liannan\}@buaa.edu.cn ~~ \{cszyluo,majing\}@comp.hkbu.edu.hk} \\
}

\begin{document}
\maketitle
\begin{abstract}
Recent advancements in large language models (LLMs) have showcased impressive code generation capabilities, primarily evaluated through language-to-code benchmarks. However, these benchmarks may not fully capture a model's code understanding abilities. We introduce \textbf{CodeJudge-Eval (CJ-Eval)}, a novel benchmark designed to assess LLMs' code understanding abilities from the perspective of code judging rather than code generation. \textbf{CJ-Eval} challenges models to determine the correctness of provided code solutions, encompassing various error types and compilation issues. By leveraging a diverse set of problems and a fine-grained judging system, \textbf{CJ-Eval} addresses the limitations of traditional benchmarks, including the potential memorization of solutions. Evaluation of 12 well-known LLMs on \textbf{CJ-Eval} reveals that even state-of-the-art models struggle, highlighting the benchmark's ability to probe deeper into models' code understanding abilities. Our codes and benchmark are available at \url{https://github.com/CodeLLM-Research/CodeJudge-Eval}.
\end{abstract}
\section{Introduction}

Recently, powerful large language models (LLMs) such as GPT-4o~\cite{GPT4}, Gemini~\cite{gemini}, and Claude~\cite{claude} have demonstrated impressive code generation capabilities. These models are being used to develop tools that assist in software development~\cite{DBLP:conf/iclr/HongZCZCWZWYLZR24,SWE-agent}. The primary method the community uses to evaluate the coding abilities of these LLMs is based on popular language-to-code benchmarks, such as HumanEval~\cite{humeval}, APPS~\cite{APPS} and MBPP~\cite{MBPP}, where LLMs are tasked with generating code based on task descriptions. If the generated code can pass the pre-designed test cases, the LLMs are considered to have successfully solved the coding tasks.

\begin{figure}[t]
    \centering
    \includegraphics[width=1\linewidth]{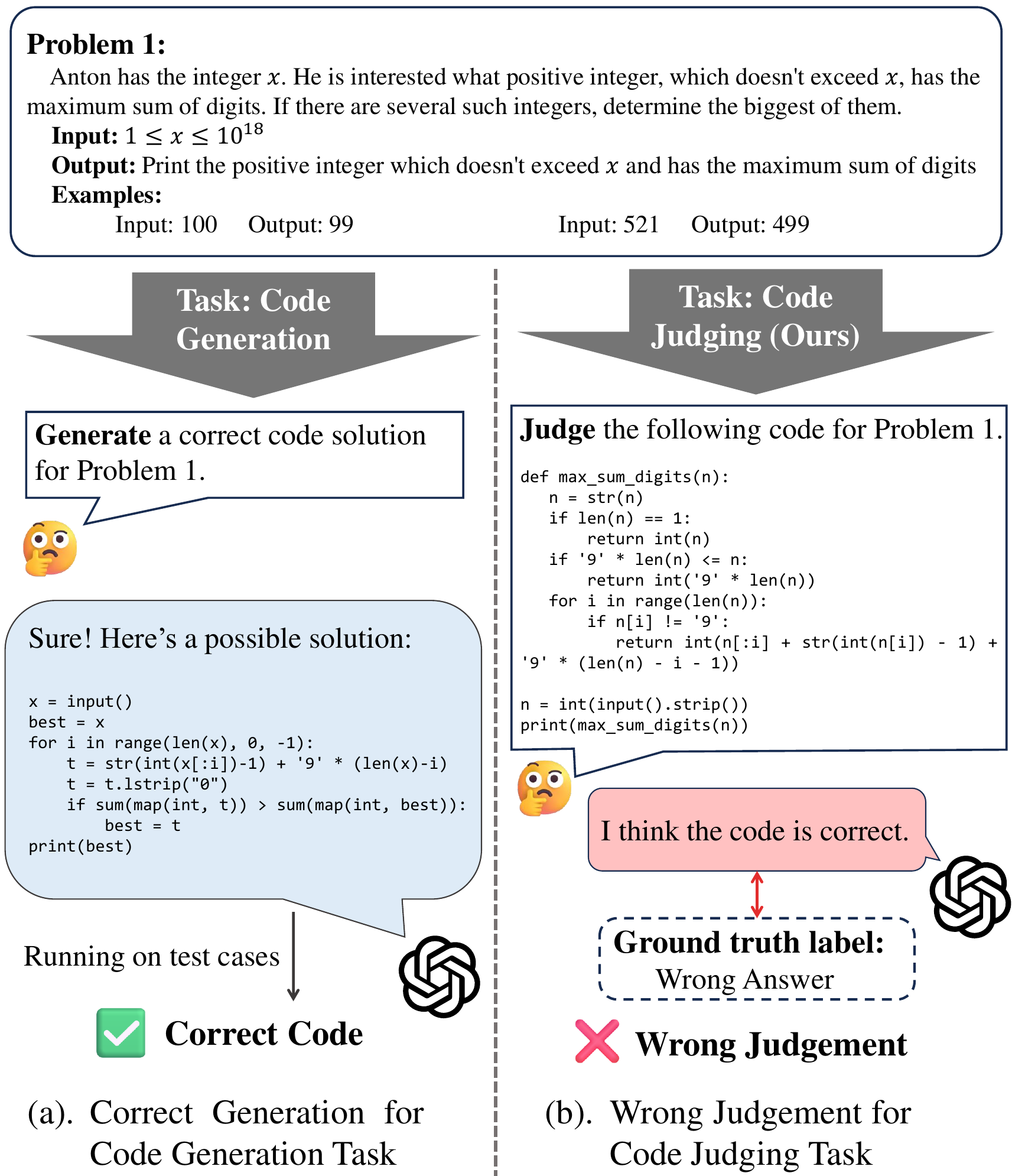}
    \caption{Comparing code generation with code judging task, we observe that a model's ability to generate correct code does not necessarily imply it can accurately judge other codes for the same problem.}
    \label{fig:intro_gap}
\end{figure}

While language-to-code benchmarks have significantly advanced the coding capabilities of LLMs, the assumption that a model’s ability to pass pre-designed test cases for a specific task equates to a full understanding of that task does not always hold true~\cite{DBLP:journals/corr/abs-2407-06153}. These test cases may not comprehensively cover all potential inputs and edge cases~\cite{humanevalp}, and concerns such as data leakage can further undermine the reliability of such evaluations~\cite{DBLP:conf/acl/DongJLJGYL24,livebench}. To overcome these challenges, we draw inspiration from modern educational theory, which suggests that if someone can accurately judge the correctness of other candidate solutions for a given task, they are likely to fully understand that task~\cite{care2012assessment}. Building on this insight, we introduce a novel benchmark, \textbf{CodeJudge-Eval (CJ-Eval)}, aimed at evaluating the code understanding abilities of LLMs by positioning them as code judges, as shown in Figure~\ref{fig:intro_gap}.

Unlike traditional approaches that require LLMs to generate code, \textbf{CJ-Eval} assesses their ability to evaluate the correctness of provided candidate solutions, determining whether they result in a correct output or errors such as Wrong Answer, Time Limit Exceeded, or other errors. Although unit tests can verify code correctness directly, our objective is to evaluate the inherent code understanding abilities of LLMs without relying on external tools, thereby reducing the need for diverse and high-quality unit tests across different coding tasks. Moreover, the LLM-as-a-Judge paradigm is already widely adopted in the general domain, as evidenced by frameworks such as MT-Bench~\cite{MT-Bench} and AlpacaEval~\cite{alpaca_eval}.

Additionally, evaluating the model using the code judging paradigm also offers new insights from a data perspective. Previous research has shown that a 7B model can memorize more knowledge than English Wikipedia~\cite{scaling_knowledge}, making it likely that the model could pass the code generation evaluation by merely memorizing one correct solution per problem. LiveCodeBench~\cite{Jain2024LiveCodeBenchHA} and LiveBench~\cite{livebench} address this issue by adding new data to the benchmark.  In contrast, our code judge evaluation assesses each code solution, and the number of code submissions is often much greater than the number of problems\footnote{For instance, Codeforces, a famous programming website, has produced only approximately 9,800 problems over the past 14 years but has $2.7\times 10^8$ solution codes.}, making it harder for the model to memorize all solutions.

To construct our \textbf{CJ-Eval} benchmark, we choose to select problems from the APPS test set, which includes 5,000 coding problems across three different difficulty levels, offering significantly more diversity than smaller benchmarks like HumanEval and MBPP. To generate candidate code solutions for each problem, we utilized 16 different LLMs, encompassing both open- and closed-source, as well as general and code-specific models. We then applied our fine-grained judging system, using a comprehensive set of test cases to obtain execution results that serve as the ground-truth judging annotations. To create a curated benchmark, we meticulously filtered the original 80,000 solutions down to 1,860 solution codes and structured the questions into a multiple-choice format.

We evaluated 12 different proprietary and open-source LLMs using our \textbf{CJ-Eval} benchmark. The results indicate that our benchmark is quite challenging. While proprietary LLMs such as GPT-4o and Claude-3.5-Sonnet outperform open-source models, their macro F1 scores peak at a modest 50 on the simplest code judge tasks. This indicates a significant gap between their performance and the highlighting room for considerable improvement. Additionally, our analysis reveals that while some models can generate correct code for certain tasks, this does not necessarily mean they can accurately judge other codes for the same tasks. This suggests that our benchmark provides a new perspective for assessing the code understanding abilities of both proprietary and open-source LLMs.
\section{Related Work}

\begin{figure*}[t]
    \centering
  \includegraphics[width=0.9\textwidth]{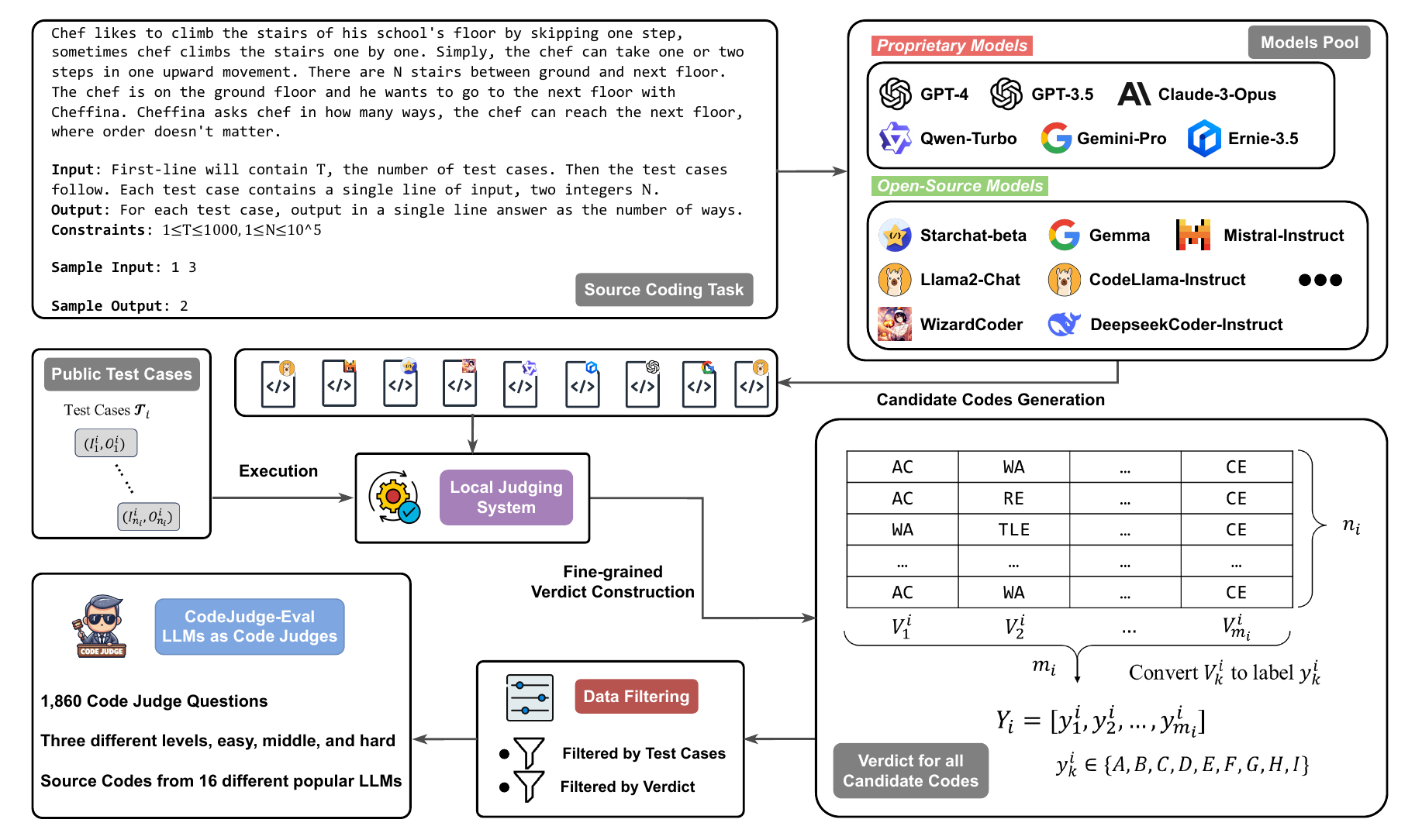}
  \caption{An overview of our pipeline for constructing the \textbf{CodeJudge-Eval} benchmark.}
  \label{fig:pipeline}
\end{figure*}

Various benchmarks are used to evaluate LLMs' coding abilities. For Python code generation on relatively simple task descriptions, HumanEval~\cite{humeval} and MBPP~\cite{MBPP} are the most popular benchmarks. EvalPlus~\cite{humanevalp} enhance HumanEval and MBPP by adding more test cases. ReCode~\cite{recode} modifies HumanEval by changing function names and docstrings to create a benchmark for code generation robustness. Extensions like HumanEval-X~\cite{CodeGeeX}, MultiPL-E~\cite{MultiPL-E}, and MBXP~\cite{MBXP} adapt HumanEval and MBPP to include programming languages beyond Python. APPS~\cite{APPS}, CodeContests~\cite{AlphaCode}, and TACO~\cite{TACO} introduce more challenging coding problems. MMCode~\cite{MMCode} extends these competition-level coding tasks with multimodal information. CodeHalu~\cite{tian2024codehalu} evaluates various hallucinations in code generation. DS-1000~\cite{DS1000}, NumpyEval~\cite{NumpyEval}, and PandasEval~\cite{pandaseval} focus on data science code generation.

Additionally, a variety of code benchmarks exist for tasks such as code translation~\cite{DBLP:conf/nips/RoziereLCL20,CodeTransOcean,DBLP:conf/eacl/AhmadCRC23}, test case generation~\cite{DBLP:journals/corr/abs-2406-04531,DBLP:journals/corr/abs-2404-13340}, code search~\cite{husain2019codesearchnet}, commit message generation~\cite{DBLP:conf/wcre/SchallCM24}, code summarization~\cite{Sun2024SourceCS}, program repair~\cite{octocoder,DBLP:journals/jss/YeMDM21,yan2023codescope}, code execution~\cite{CRUXEval}, and repository-level code generation~\cite{RepoBench,SWE-bench}. However, these benchmarks predominantly focus on code generation based on given requirements and often rely on high-quality test cases to evaluate the correctness of generated code, which can be susceptible to data leakage issues~\cite{DBLP:conf/acl/DongJLJGYL24}.

In contrast, our \textbf{CJ-Eval} evaluates LLMs' code understanding abilities from the perspective of LLMs acting as code judges. This approach does not depend on external unit testing; instead, it requires LLMs to evaluate various solutions for the same task, increasing the difficulty of memorizing all possible solutions. Similar efforts include ICE-Score~\cite{DBLP:conf/eacl/Zhuo24}, which introduces a metric for evaluating the usefulness of code generated by LLMs, and the study by~\citet{DBLP:conf/acl/GuLJOLSS24}, which examines LLMs' challenges in understanding the nuances of their own incorrect generations. However, our work uniquely introduces a benchmark focused specifically on code correctness judgment, evaluating whether LLMs can assess code not only generated by themselves but also by other LLMs.
\section{CodeJudge-Eval}
\label{sec:CodeJudgeEval}

\subsection{Overview}
\label{sec:overview}

As shown in Figure~\ref{fig:pipeline}, we introduce the construction pipeline of our \textbf{CJ-Eval} benchmark. It consists of $N$ problems, denoted as $P_1, \ldots, P_N$. A problem $P_i$ with $n_i$ test cases and $m_i$ solution codes can be formatted as
\begin{equation}
  \label{eq:problem_def}
  P_i = (S_i, \mathcal{T}_i, \mathcal{C}_i, Y_i, [V^i_1, \ldots , V^i_{m_i}]).
\end{equation}
\begin{equation}
  \label{eq:test_cases_def}
  \mathcal{T}_i = \{(I_1^i, O_1^i), \ldots, (I_{n_i}^i, O_{n_i}^i)\}
\end{equation}
\begin{equation}
  \label{eq:codss_def}
\mathcal{C}_i = \{c^i_1, \ldots, c^i_{m_i}\}.
\end{equation}
$P_i$ is composed of:
\begin{itemize}
    \setlength{\itemsep}{0.1em}
    \item A problem statement $S_i$ in text form. 
    \item A set of test cases $\mathcal{T}_i$, where $I_j^i$ and $O_j^i$ denote the input and output of the $j$-th test case, respectively.
    \item A set of solution codes $\mathcal{C}_i$, which consists of $m_i$ solution codes generated by models for problem $P_i$.
    \item A list of verdicts\footnote{Some programming websites (e.g., Codeforces) use the term ``verdict'' to indicate the result of code execution. We adopt this term in our work.} $[V^i_1, \ldots , V^i_{m_i}]$. For the $k$-th solution code $c_k^i$, we define $V^i_k = [v_1, \ldots, v_{n_i}]$ to represent the results of running solution code $c_k^i$ on all $n_i$ test cases $\mathcal{T}_i$, in the form of a list of verdicts. Note that for a given code, the verdicts are in the form of a list indicating its results on $n_i$ test cases. For example, when $n_i = 4$, a possible verdicts for $c_k^i$ could be $V^i_k = [\mathrm{AC}, \mathrm{WA}, \mathrm{TLE}, \mathrm{AC}]$.
    \item A list of labels $Y_i$. The list of labels (or choices) for $m_i$ solution codes are $Y_i = [y_1^i, \ldots, y_{m_i}^i]$, which is derived from \([V_1^i, \ldots, V_{m_i}^i]\), where $V_k^i$ determines the label $y_k^i$. $Y_i$ represents the label we use for the final evaluation, and each code has only one label. 
\end{itemize}

\subsection{Dataset Construction}
\label{sec:dataset_construction}


Following Equation~\ref{eq:problem_def}, we introduce the data sources used to obtain $S_i$ and $\mathcal{T}_i$ in Section~\ref{sec:data_source}. In Section~\ref{sec:code_generation}, we explain how we generated the solution codes $\mathcal{C}_i$. Finally, in Section~\ref{sec:verdict_construction}, we detail the rules for constructing the labels $Y_i$.

\subsubsection{Data Source}
\label{sec:data_source}

We used the test set from the APPS~\cite{APPS} dataset as our data source, given its challenging nature and the the abundance of test cases. The APPS test set consists of 5,000 programming problems sourced from websites such as Codeforces, LeetCode, and Kattis. The problems are categorized into three levels of difficulty: introductory, interview, and competition. Each problem includes a problem statement $\mathcal{S}_i$ and multiple test cases $\mathcal{T}_i$. Each $\mathcal{S}_i$ contains a problem description along with several input-output examples. It is important to note that there is at least one test case not included in $\mathcal{S}_i$, meaning there are some hidden test cases. We used these 5,000 problems as our raw data.

\subsubsection{Code Generation}
\label{sec:code_generation}
To generate solution codes $\mathcal{C}_i$, we select 16 representative LLMs which are capable of code generation. To ensure diversity in the generated code, we consider three different categories of LLMs: proprietary general-purpose LLMs, including GPT-4~\cite{GPT4}, GPT-3.5~\cite{GPT3}, Claude-3-Opus~\cite{claude}, Gemini-1.0-pro~\cite{gemini}, Ernie-3.5~\cite{BaiduRes10:online}, and Qwen-turbo~\cite{qwen}; open-source generalist LLMs, including Starchat~\cite{li2023starcoder}, Gemma~\cite{gemma}, Mistral-Instruct~\cite{mistral}, Llama2-chat~\cite{llama2}, and ChatGLM3~\cite{chatglm3}; and open-source code LLMs, including CodeLlama-Instruct~\cite{codellama}, Magicoder-S-DS~\cite{magiccoder}, WizardCoder-Python~\cite{wizardcoder}, and DeepseekCoder-Instruct~\cite{deepseekcoder}.

Our prompt included the problem statement $S_i$ and a request to generate corresponding Python code. In most cases, we extract the code by identifying the \verb|```python(.*?)```| regular expression from the model outputs.

\subsubsection{Fine-grained Verdict Construction}
\label{sec:verdict_construction}
After obtaining $S_i, \mathcal{T}_i, \mathcal{C}_i$, we need to evaluate a code $c^i_k$ on the $n_i$ test cases $\mathcal{T}_i$ to generate a list of verdicts $V^i_k$. To achieve this, we re-implemented a local fine-grained judging system to accurately evaluate $[V^i_1, \ldots , V^i_{m_i}]$. Unlike the original judging system of the APPS dataset, which primarily focuses on whether the code is correct without differentiating specific errors in each test case, our judging system executes each test case separately and precisely identifies the specific errors occurring in each one.

\begin{table}[t]
\small
\centering
\begin{adjustbox}{width=0.47\textwidth}
\renewcommand{\arraystretch}{1.1}
\begin{tabular}{ccc}
\toprule
 & \multirow{2}{*}{\textbf{Labels}} & \textbf{Demonstration and}                     \\ 
 & & \textbf{Example $V_k^i$} \\
\midrule
\midrule
 & \multirow{2}{*}{A}      & AC on all test cases               \\
 &       & [AC, AC, AC, AC]                   \\ \hline
 & \multirow{2}{*}{B}      & The code fails to compile          \\
 &                         & [CE, CE, CE, CE]                   \\ \hline
 & \multirow{2}{*}{C}      & Not A or B, having only WA         \\
 &                         & [AC, WA, WA, AC]                   \\ \hline
 & \multirow{2}{*}{D}      & Not A or B, having only RE         \\
 &                         & [AC, AC, AC, RE]                   \\ \hline
 & \multirow{2}{*}{E}      & Not A or B, having WA and RE       \\
 &                         & [AC, WA, RE, WA]                   \\ \hline
 & \multirow{2}{*}{F}      & Not A or B, having only TLE        \\
 &                         & [AC, AC, TLE, TLE]                 \\ \hline
 & \multirow{2}{*}{G}      & Not A or B, having WA and TLE      \\
 &                         & [WA, WA, TLE, WA]                  \\ \hline
 & \multirow{2}{*}{H}      & Not A or B, having RE and TLE      \\
 &                         & [AC, RE, RE, TLE]                  \\ \hline
 & \multirow{2}{*}{I}      & Not A or B, having WA, RE, and TLE \\
  &                        & [WA, RE, RE, TLE]                 \\
\bottomrule
\end{tabular}
\end{adjustbox}
\caption{Demonstration and example for each label. The example illustrates the list of verdicts $V_k^i$ obtained by evaluating the code $c_k^i$ on a set of test cases $\mathcal{T}_i$ with a length of $n_i = 4$.}
\label{tab:choices}
\end{table}

Following the verdict design of well-known programming websites such as Codeforces and LeetCode, we considered five types of verdicts:

\begin{itemize}
\setlength\itemsep{-0.2em}
\item  \textbf{Compilation Error\footnote{Technically, Python is an interpreted language and there is no actual compilation process. However, Python performs a syntax check before running the code, including verifying matching parentheses and correct indentation. If an error is found, exceptions like \texttt{SyntaxError} or \texttt{IndentationError}, etc., are raised. This type of error is referred to as a compilation error on Codeforces, hence we adopt the same term.} (CE)}: The code is flagged for syntax errors before execution.

\item \textbf{Runtime Error (RE)}: The code throws an exception during execution, such as an \verb|IndexError| when accessing an out-of-bounds index.

\item \textbf{Time Limit Exceeded (TLE)}: The code exceeds the time limit (2 seconds) for a single test case, typically due to suboptimal time complexity.

\item \textbf{Wrong Answer (WA)}: The code executes within the time limit but produces an incorrect output.

\item \textbf{Accepted (AC)}: The code executes within the time limit and produces the correct output.
\end{itemize}

\subsubsection{Design of Labels}
\label{sec:labels}

With $V_k^i$ evaluated, we can determine the result of $c_k^i $ for the problem, denoted as the label (or choice) $y_k^i$. Given the five possible verdicts, there are $2^5 = 32$ possible labels based on whether each verdict occurs. However, some of these cases are impossible. For instance, a code with a compilation error cannot produce any other result (since it always fails to compile). Additionally, it is unreasonable to require the model to evaluate whether there is an AC test case. Therefore, we ultimately define nine labels based on the possible list of verdicts \( V_k^i \), as shown in Table~\ref{tab:choices}.

We refer to the labels in Table~\ref{tab:choices} as the hard setting because they require the model to correctly analyze all possible error types in the code. To comprehensively evaluate models of varying levels, we additionally introduce a middle setting and an easy setting. The middle setting has six labels, where original labels \verb|EGHI| are grouped into a single label, ``Not A or B, having at least two types of errors''. In the easy setting, there are only three labels, with original labels C through I grouped into a single option, ``Not A or B, having errors''.

\subsection{Data Filtering}
\label{sec:filter}
After constructing our dataset, we obtain a dataset comprising $N$ = 5,000 problems, each with $m_i$ = 16 solution codes (this full dataset will also be released). To ensure a curated dataset, we further filter the problems and solution codes.

\begin{figure}[t]
  \includegraphics[width=\columnwidth]{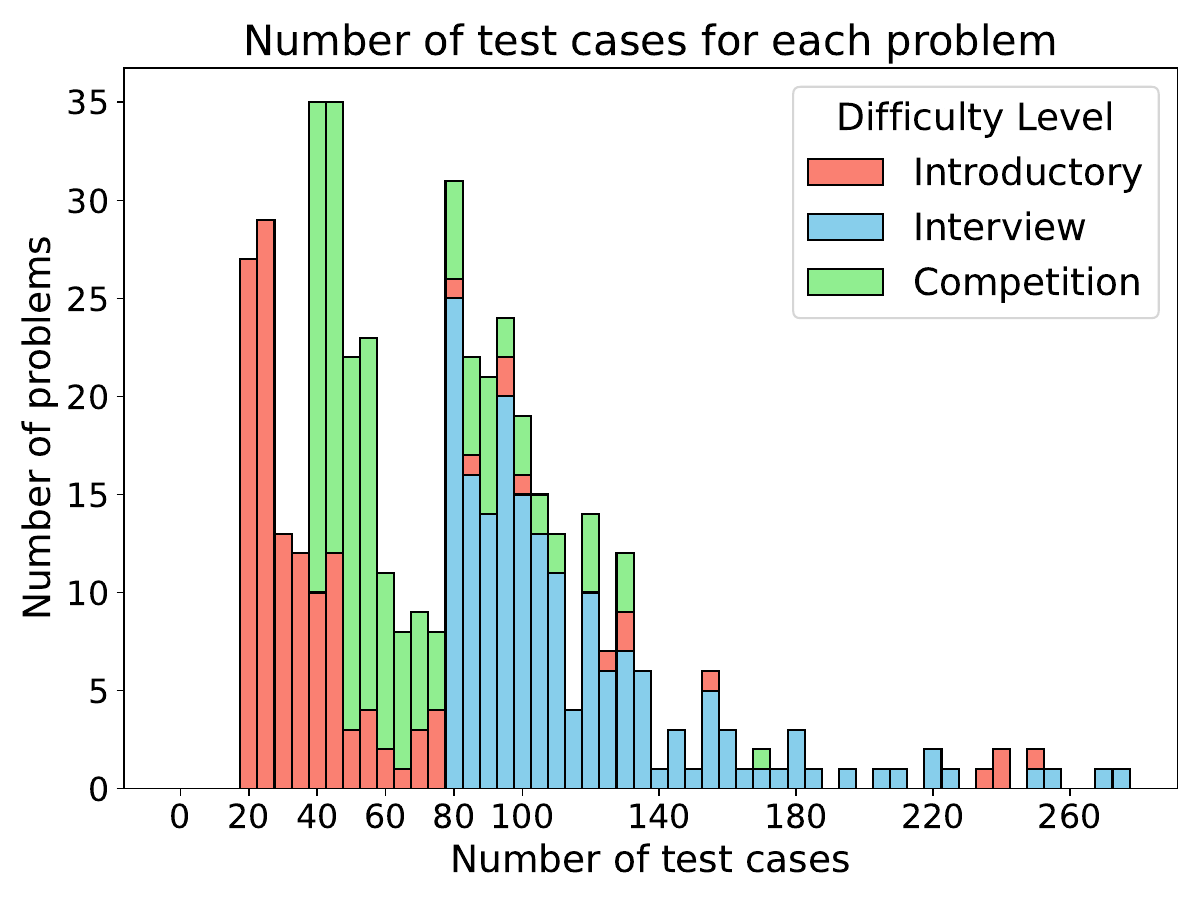}
  \caption{A stacked histogram on the number of test cases in the filtered problems. Different filtering thresholds are applied based on different difficulty.}
  \label{fig:hist}
\end{figure}

\subsubsection{Filter Problems by Test Case}
Since our benchmark requires detecting all possible error types, problems with a small number of test cases may lead to inaccurate labels. For example, if a problem has very few test cases, a program that achieves AC on it might encounter WA, TLE, or other errors when tested against more rigorous cases. 

To ensure the accuracy of the labels, we applied a threshold to filter problems based on the number of test cases. Specifically, we exclude problems with no more than 20, 80, and 40 test cases for the introductory, interview, and competition levels, respectively. After this filtering process, we obtain a total of 457 problems, including 133 at the introductory level, 178 at the interview level, and 146 at the competition level. We visualize the distribution of the number of test cases for problems of each difficulty level in Figure~\ref{fig:hist}.

\subsubsection{Filter Solution Codes by Verdict}

So far, we have obtained a dataset subset with $N$ = 457 and $m_i$ = 16. To further refine the dataset, we considered redundancy in verdicts for the same problem. For each problem, we retain only one solution code per verdict (A to I). The selection of which solution code to retain is made randomly. The detailed filtering algorithm can be found in Appendix~\ref{sec:algo_filter}.

After filtering, there are a total of 1,860 solution codes (i.e., $\Sigma m_i$ = 1,860), with an average of 4.1 codes per problem. For the easy and middle settings introduced in Section~\ref{sec:labels}, we still use these 1,860 data points, but their labels have been modified according to the rules above. The statistical information regarding the models and verdicts corresponding to these solution codes is provided in Appendix~\ref{sec:stat_easy_middle}.
\begin{table*}[h!]
    \centering
    \begin{tabular}{lccccccc}
        \toprule
        \multirow{2}{*}{\textbf{Model}} & \multirow{2}{*}{\textbf{Size}} & \multicolumn{2}{c}{\textbf{Easy}} & \multicolumn{2}{c}{\textbf{Middle}} & \multicolumn{2}{c}{\textbf{Hard}}\\
        & & \textbf{Acc} & \textbf{F1} & \textbf{Acc} & \textbf{F1} & \textbf{Acc} & \textbf{F1}\\
        \midrule
        \midrule
        \rowcolor{gray!40}
        \multicolumn{8}{c}{\textit{Simple Strategies}}\\
        \textbf{Random} & - & 33.76 & 25.65 & 16.29 & \textbf{15.08} & 12.31 & \textbf{10.75} \\
        \textbf{Always AC} & - & 9.57 & 5.82 & 9.57 & 2.91 & 9.96 & 2.26 \\
        \textbf{Always Most Frequent Choice} & - & \textbf{81.18} & \textbf{29.87} & \textbf{31.34} & 7.95 & \textbf{25.52} & 5.08 \\
        \midrule
        \rowcolor{pink!50}
        \multicolumn{8}{c}{\textit{Proprietary Models}}\\
        \Openaiemoji{}~\textbf{GPT-4o} & - & \textbf{84.30} & 38.16 & \textbf{31.56} & 20.67 & 30.75 & 13.61\\
        \midrule
        \Anthropicemoji{}~\textbf{Claude-3.5-Sonnet} & - & 80.11 & \textbf{50.83} & 31.18 & \textbf{27.02} & \textbf{30.86} & \textbf{19.05}\\
        \midrule
        \Googleemoji{}~\textbf{Gemini-1.5-Pro} & - & 80.38 & 33.91 & 31.29 & 22.65 & 28.39 & 15.76\\
        \midrule
        \Openaiemoji{}~\textbf{GPT-3.5-Turbo} & - & 38.06 & 18.68 & 16.24 & 10.31 & 12.63 & 5.83\\
        \midrule
        \rowcolor{green!30}
        \multicolumn{8}{c}{\textit{Open-Source Generalist Models}}\\
        \Mistralemoji{}~\textbf{Mistral-Nemo-Instruct} & 12B & 9.62 & 4.55 & 9.46 & 2.52 & 9.52 & 1.76\\
        \midrule
        \Googleemoji{}~\textbf{Gemma2-IT} & 9B & \textbf{57.04} & \textbf{19.80} & \textbf{19.14} & 9.30 & \textbf{18.87} & \textbf{9.17}\\
        \midrule
        \Metaemoji{}~\textbf{Llama-3.1-Instruct} & 8B & 13.01 & 11.81 & 10.11 & \textbf{9.74} & 9.03 & 7.69\\
        \midrule
        \Qwenemoji{}~\textbf{Qwen2-Instruct} & 7B & 21.88 & 14.51 & 16.99 & 7.56 & 9.89 & 3.51\\
        \midrule
        \rowcolor{green!30}
        \multicolumn{8}{c}{\textit{Open-Source Code Models}}\\
        \Qwenemoji{}~\textbf{CodeQwen1.5-Chat} & 7B & 15.05 & 13.03 & \textbf{9.89} & \textbf{3.95} & \textbf{10.00} & \textbf{3.37}\\
        \midrule
        \Metaemoji{}~\textbf{CodeLlama-Instruct} & 7B & \textbf{59.84} & \textbf{21.15} & 5.16 & 3.78 & 5.48 & 3.13\\
        \midrule
        \Googleemoji{}~\textbf{CodeGemma-IT} & 7B & 16.40 & 10.39 & 5.48 & 3.69 & 5.59 & 3.17\\
        \midrule
        \Deepseekemoji{}~\textbf{DeepseekCoder-Instruct} & 6.7B & 10.38 & 7.28 & 9.73 & 2.80 & 9.68 & 1.97\\
        \bottomrule
    \end{tabular}
    \caption{Zero-shot accuracy and macro F1 scores on the \textbf{CJ-Eval} benchmark, evaluating four method types across three difficulty levels.}
    \label{tab:main_results}
\end{table*}
\section{Experiment}

\subsection{Experimental Setup}

\subsubsection{Evaluated Methods}

For all methods, we use a temperature of 0.0. The extraction of the choice is performed using several regular expressions. Specifically, if we extract nothing, we consider the model fails to generate an answer. A discussion and statistics on failure cases can be found in Appendix~\ref{sec:fail_case}.

\paragraph{Baselines} To better assess the difficulty of our benchmark, we implement three simple rule-based strategies. ``\textbf{Random}'' randomly selects one possible choice for each problem. ``\textbf{Always AC}'' assumes that the model responds with AC for all solution codes, while ``\textbf{Always Most Frequent Choice}'' assumes that the model always selects the most frequently occurring answer in the current setting. For distributions across different difficulty settings, please refer to Appendix~\ref{sec:stat_easy_middle}.


\paragraph{Proprietary Models.} We evaluate four widely used and SOTA proprietary models~\cite{GPT4,claude,gemini}. The specific versions of the four models evaluated are \verb|gpt-4o-2024-08-06| for GPT-4o, \verb|gpt-3.5-turbo-0125| for GPT-3.5-Turbo, \verb|claude-3-5-sonnet-20240620| for Claude, and \verb|gemini-1.5-pro| for Gemini Pro.

\paragraph{Open-Source Models.} Among open-source models, we considered two types of code-capable models. The first type is open-source generalist models, which are trained for general purposes and have code capabilities as one of their skills. The second type is open-source code LLMs, which are specifically trained for code-related tasks. Considering that our tasks involve both code generation and task requirement understanding, we chose the Instruct (or Chat) versions of these models. In the Appendix~\ref{app:links}, we provide the download links for all open-source LLMs. 

\subsubsection{Metrics}

We use accuracy and macro F1 as our evaluation metrics. Accuracy intuitively reflects the proportion of correctly answered questions out of 1,860 problems. However, considering the class imbalance issue, particularly in the easy setting where class \verb|C| accounts for around 81\% (see Appendix~\ref{sec:stat_easy_middle} for details), we introduce macro F1 to accurately assess the model's overall performance across all classes. Macro F1 is calculated by averaging the F1 scores of each class, defined as
\begin{equation}
  \label{eq:macro_f1}
  \mathrm{F1}=\frac{1}{n} \sum_c \frac{2P_cR_c}{P_c+R_c},
\end{equation}
where $c$ represents the class, and $P, R$ denote precision and recall, respectively. We recommend using macro F1 as the primary metric.

\subsection{Zero-Shot Evaluation}

In Table~\ref{tab:main_results}, we evaluate the zero-shot performance of four different types of methods across three difficulty levels on our \textbf{CJ-Eval} Benchmark. The performance of all LLMs is suboptimal, highlighting the challenging nature of our benchmark.

\paragraph{On Simple Strategies.} We first focus on simple strategies, as they serve as baselines for comparison. The ``Always AC'' strategy performs poorly across both metrics. The ``Always Most Frequent Choice'' strategy achieves the highest accuracy; however, its F1 score is either lower than or only slightly better than that of the ``Random'' strategy. This indicates that although simple strategies can achieve high accuracy due to class imbalance, none of them significantly outperform in terms of F1.

\paragraph{Comparing Propriety Models.} Proprietary models achieve the best overall performance among all methods.
Except for GPT-3.5, the performance of three proprietary models is significantly superior to that of open-source models. Moreover, their macro-F1 scores are higher than those of simple strategies, indicating that their judgments are not based on some simple tricks.

\paragraph{Comparing Open-Source Models.} Overall, open-source models perform poorly on our benchmark. Almost all models have lower macro F1 scores compared to the ``Random'' strategy. Furthermore, while open-source code LLMs like DeepseekCoder have shown comparable performance~\cite{deepseekcoder} with GPT-3.5-turbo on HumanEval~\cite{humeval} and MBPP~\cite{MBPP}, they perform much worse on our benchmark compared to GPT-3.5-turbo. This suggests that our evaluation offers a new perspective for investigating the potential gap between open-source and proprietary models.

\subsection{Analysis}

\begin{figure}[t]
    \centering
    \includegraphics[width=0.8\linewidth]{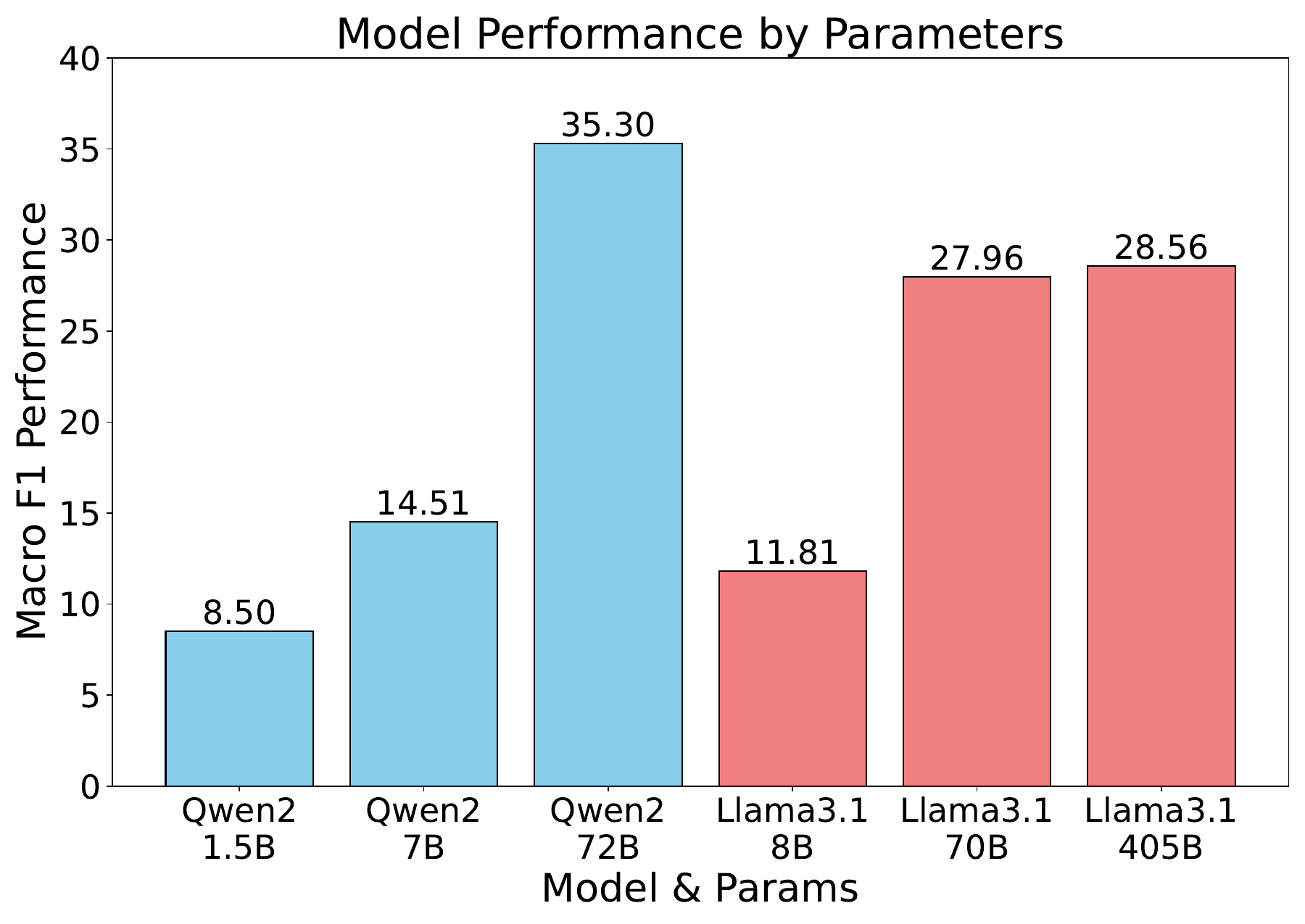}
    \caption{Scaling the number of models' parameters on our \textbf{CJ-Eval} Easy.}
    \label{fig:ana_params}
\end{figure}

\paragraph{Do More Parameters Help?}

Figure~\ref{fig:ana_params} presents a comparison of performance when scaling two well-known open-source LLMs, Qwen2 and Llama-3.1, across various parameter sizes. The results indicate that increasing the size of LLMs can significantly enhance code judging performance. Notably, Qwen2-72B achieves performance levels comparable to those of GPT-4o. However, continuously scaling the parameter size of Llama-3.1 from 70B to 405B yields only marginal improvements, suggesting that merely increasing the number of parameters does not necessarily lead to substantial gains in code judging performance.

\begin{figure}[t]
    \centering
    \includegraphics[width=0.8\linewidth]{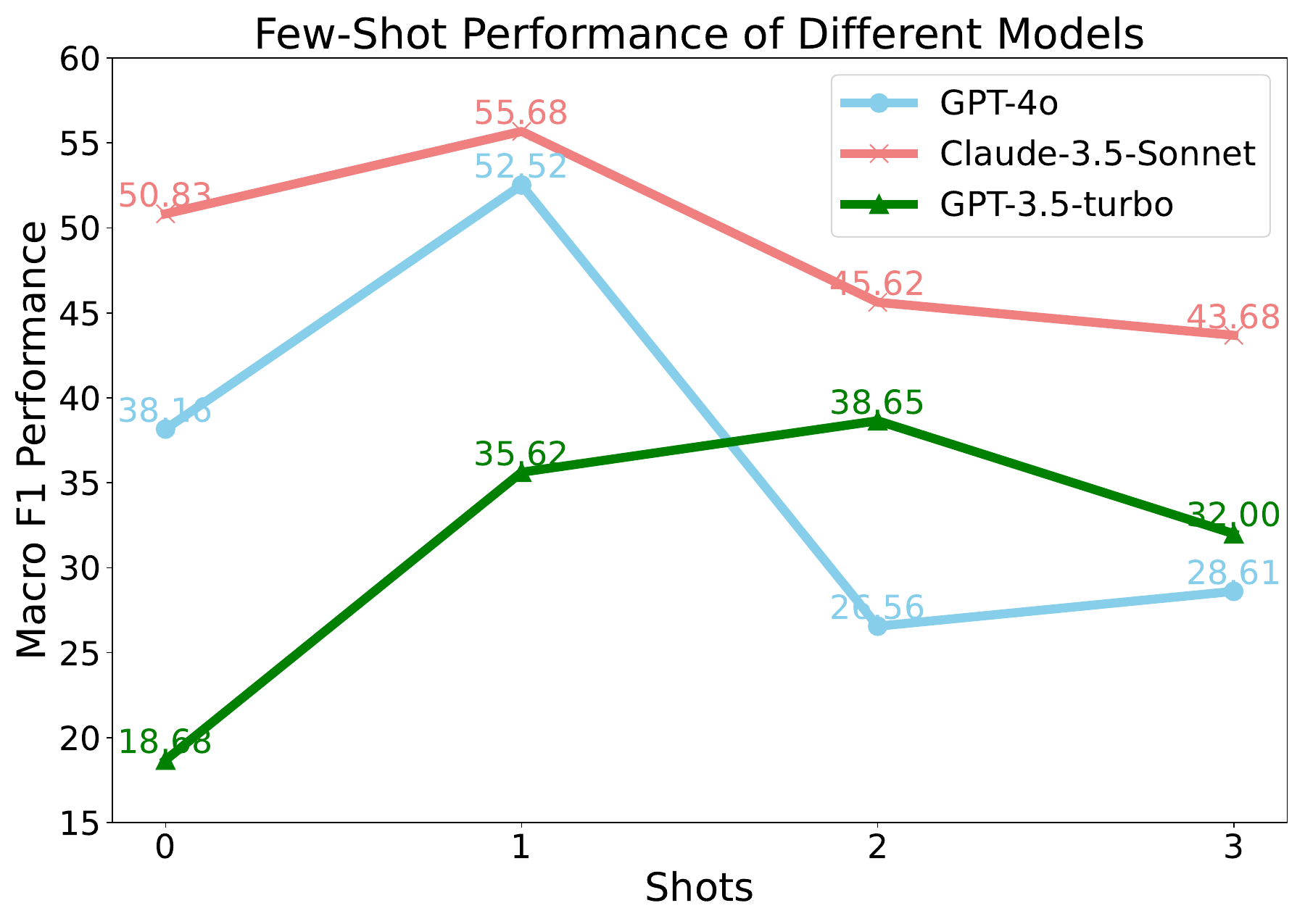}
    \caption{Comparative few-shot F1 scores of different models on our \textbf{CJ-Eval} easy.}
    \label{fig:ana_few_shot}
\end{figure}

\paragraph{Do Few-Shot Examples Help?}

As illustrated in Figure~\ref{fig:ana_few_shot}, few-shot examples can offer some benefits in enhancing model performance on this benchmark. With one-shot examples, there is a significant increase in the performance of all models: +14.36 for GPT-4o, +4.85 for Claude-3.5-Sonnet, and +16.94 for GPT-3.5-Turbo. However, performance tends to decline substantially as the number of shots increases. A plausible explanation for this phenomenon is that longer prompts, which result from additional shots, might detrimentally affect the reasoning capabilities of LLMs. More few-shot results are provided in Appendix~\ref{app:all_results}.

\begin{figure}[t]
    \centering
    \includegraphics[width=0.9\linewidth]{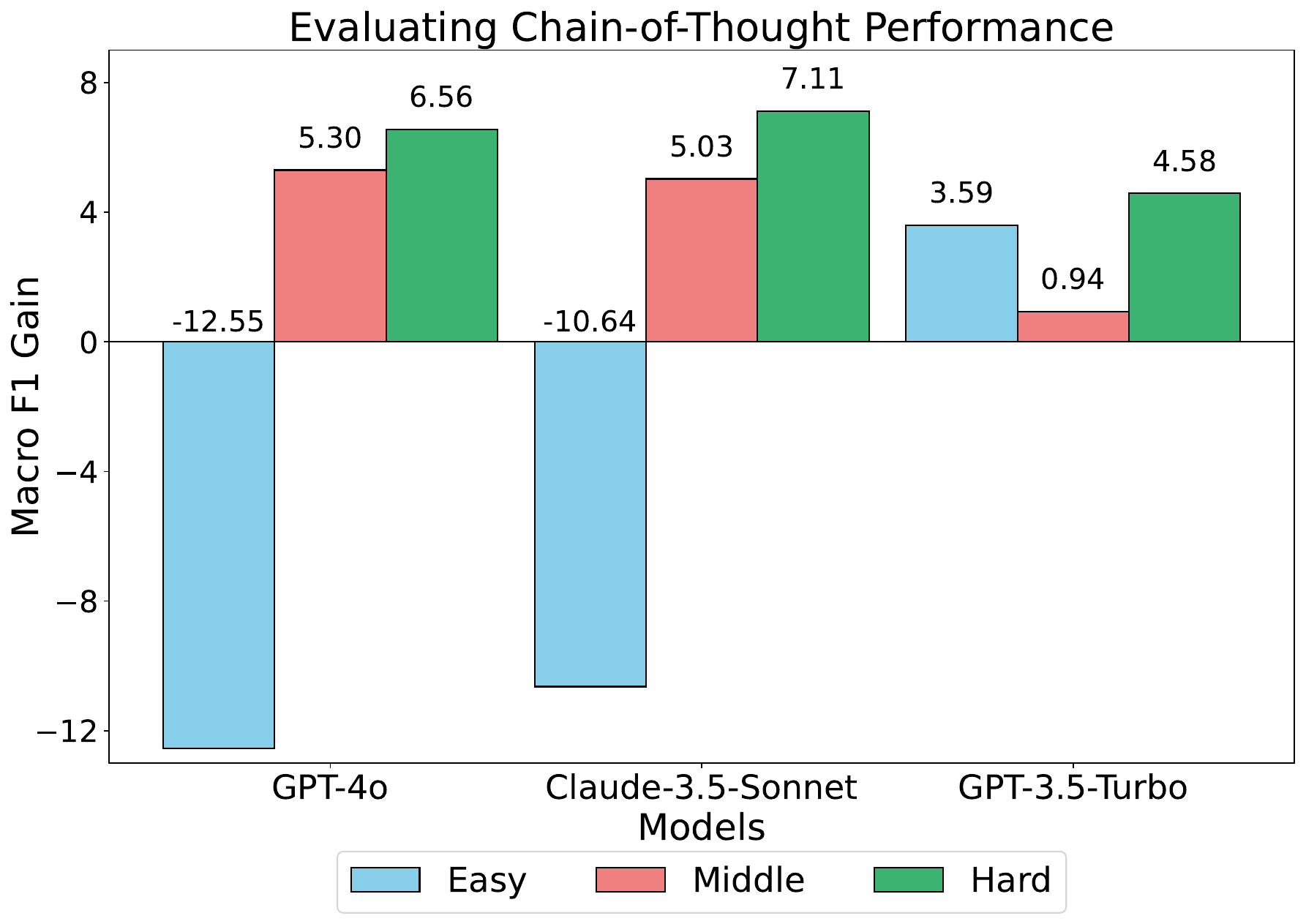}
    \caption{Evaluating Chain-of-Thought performance. We show the Macro F1 gain brought by 1-shot CoT example compared to vanilla 1-shot example.}
    \label{fig:CoT}
\end{figure}

\paragraph{Does Chain-of-Thought Example Help?}

We further design a 1-shot Chain-of-Thought~\cite{cot} (CoT) example, which is presented in the Appendix~\ref{sec:prompt}. For a fair comparison, we compared the performance of the 1-shot CoT with that of a vanilla 1-shot example. Figure~\ref{fig:CoT} illustrates the macro F1 gain achieved by the CoT example. It can be seen that the CoT example provides significant guidance in the middle hard setting, but may lead to a performance decrease in the easy setting. This decrease may be due to the fact that in the easy setting, the task only requires determining whether the code is correct, without needing the detailed analysis provided by the CoT example. The complete results can be found in the Appendix~\ref{app:all_results}.

\begin{figure}[t]
    \centering
    \includegraphics[width=0.9\linewidth]{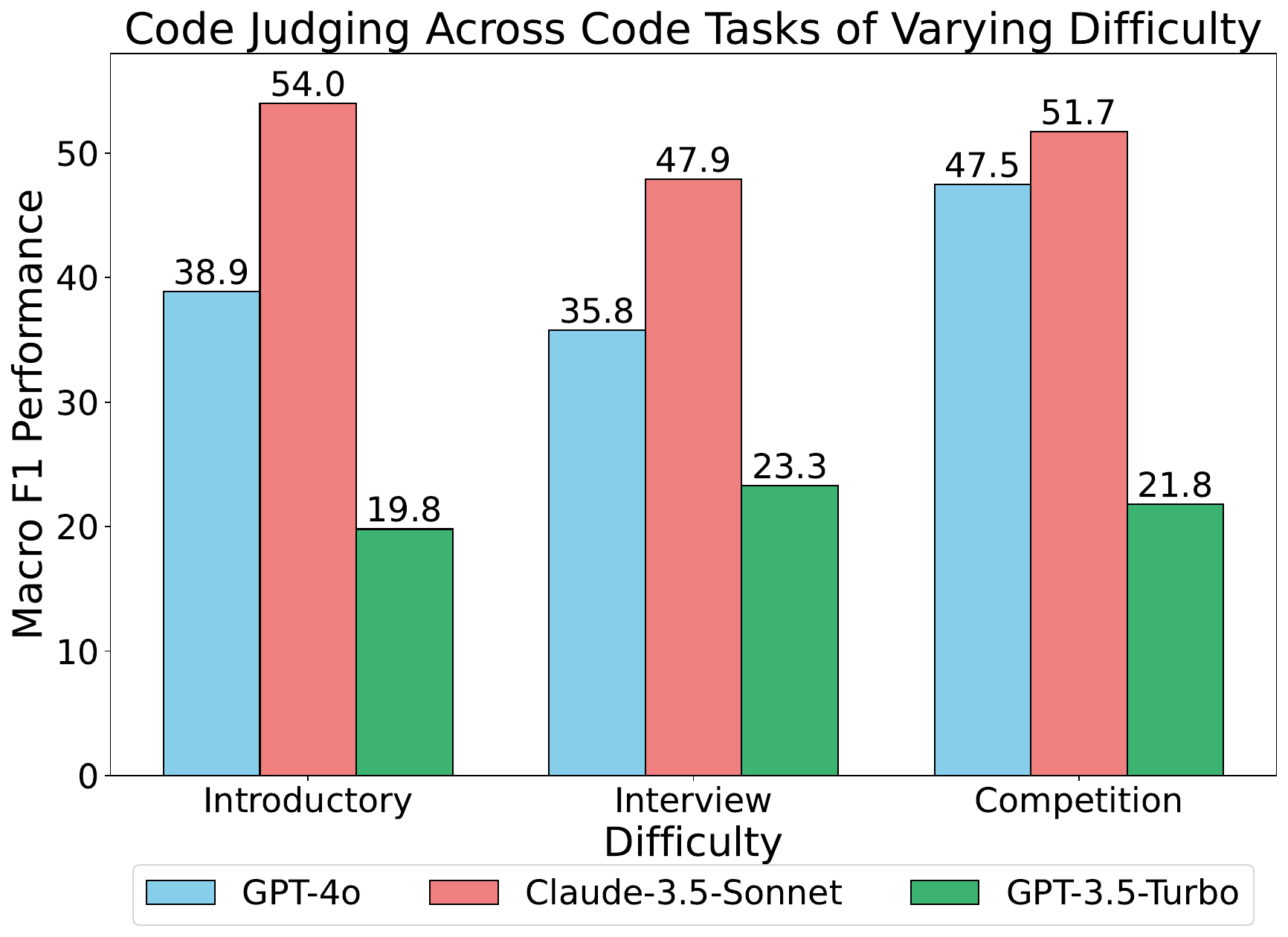}
    \caption{Analyzing whether the easier coding tasks are easier to judge.}
    \label{fig:diff_judge}
\end{figure}

\paragraph{Are Easier Coding Tasks Easier to Judge?}

\begin{figure}[t]
    \centering
    \includegraphics[width=0.9\linewidth]{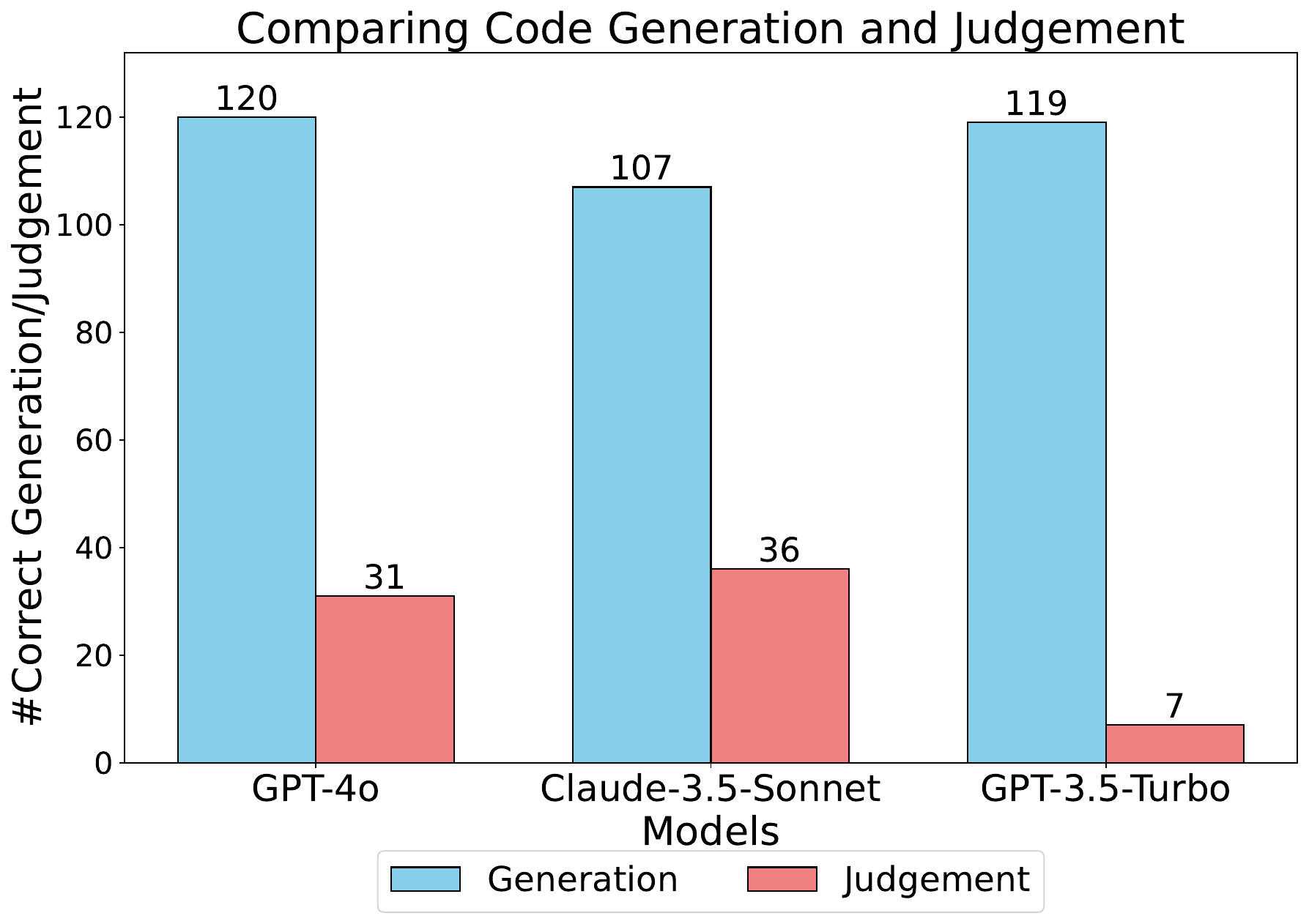}
    \caption{Analyzing whether the ability to generate correct code for a task guarantees the ability to judge the correctness of other codes for the same tasks.}
    \label{fig:gap}
\end{figure}

The source for the coding tasks in our CJ-Eval is derived from the APPS test set, where tasks are categorized into three difficulty levels: introductory, interview, and competition, in ascending order of complexity. As depicted in Figure~\ref{fig:diff_judge}, it is evident that easier tasks do not necessarily yield higher macro F1 judging scores, indicating that the evaluated capabilities in code judging tasks differ from those in code generation tasks. 

\paragraph{Does Accurate Generation Ensure Accurate Judgment?}

In Figure~\ref{fig:gap}, we examine whether LLMs' ability to generate correct code translates to its ability to assess the correctness of candidate solutions. We first select tasks for which the LLMs can generate correct code. Next, we evaluate how often these LLMs can accurately judge solutions produced by other models. The results highlight a notable gap between generation and judgment capabilities, indicating that generating correct code does not guarantee the ability to assess code correctly. Consequently, our benchmark offers a distinct perspective on LLMs' coding abilities.
\section{Conclusion}

In this work, we present \textbf{CodeJudge-Eval}, a novel benchmark designed to evaluate LLMs' code understanding capabilities by assessing their performance as code judges. We tested 12 popular LLMs, both proprietary and open-source, on our benchmark. The results demonstrate the benchmark’s difficulty, with open-source models often performing worse than random guessing. Moreover, our analysis shows that a model’s ability to generate correct code does not necessarily imply it can accurately evaluate other solutions for the same task.

\section*{Limitation}

Our benchmark has room for enhancement in several aspects:
\begin{itemize}
    \item While the experimental results of \textbf{CJ-Eval} offer a novel perspective for assessing the code comprehension capabilities of LLMs, it is not intended as a replacement for existing language-to-code benchmarks. As our benchmark does not evaluate code generation, a more comprehensive approach would involve integrating our benchmark with language-to-code evaluations to more effectively assess the code understanding performance of LLMs.
    \item In the introduction, we discussed that our benchmark incorporates multiple candidate solutions for each coding task, making it more challenging for LLMs to memorize all possible solutions and thus circumvent our evaluation. However, we acknowledge that this approach does not fully mitigate the risk of intentional attempts to train LLMs to memorize all candidate solutions.
    \item The design of our benchmark is specifically tailored to coding domains, thereby limiting its applicability across broader, more general domains. Its foundation in code judging principles poses significant challenges in adapting the methodology for non-coding contexts.
\end{itemize}


\bibliography{coling}

\appendix
\begin{algorithm*}
\caption{Filter codes for $i$-th problem}
\label{alg:process}
\begin{algorithmic}[1]
\REQUIRE $P_i = (S_i, \mathcal{T}_i, \mathcal{C}_i, [V^i_1, \ldots , V^i_{m_i}], Y_i)$
\ENSURE $P_i$ after filtering solution codes
\STATE Initialize set of verdicts $R_x = \emptyset$ \COMMENT{Record existing verdicts for the $i$-th problem}
\STATE Initialize set of index $\mathcal{U} = \{1, \dots, m_i\}$ \COMMENT{$m_i=16$ in this algorithm}
\STATE Initialize set of index to be deleted $\mathcal{D} = \emptyset$
\WHILE{$\mathcal{U}$ is not empty}
    \STATE Randomly select an index $k$ from $\mathcal{U}$ and delete $k$ from $\mathcal{U}$
    \IF[Recall that $y_k^i$ is a letter from A to I]{$y_k^i\notin R_x$} 
        \STATE Add $k$ to $\mathcal{D}$ \COMMENT{Code $c_k^i$ is retained}
    \ELSE
        \STATE Add $y_k^i$ to $R_x$ \COMMENT{Code $c_k^i$ is discarded}
    \ENDIF
\ENDWHILE
\FOR{deleted index $k$ in $\mathcal{D}$} 
    \STATE Tag \(c_k^i\), \(V^i_k\), and \(y^i_k\) as discarded data
\ENDFOR
\STATE Remove the discarded data in $\mathcal{C}_i, [V^i_1, \ldots , V^i_{m_i}]$, and $Y_i$ simultaneously
\end{algorithmic}
\end{algorithm*}

\begin{table*}[ht]
\centering
\renewcommand{\arraystretch}{1.1}
\begin{tabular}{lcccccccccc}
\toprule
\multirow{2}{*}{\textbf{Model}} & \multicolumn{9}{c}{\textbf{Choice}}                      & \multirow{2}{*}{\textbf{SUM}} \\
\cline{2-10}
& A   & B   & C   & D   & E   & F   & G   & H  & I  &\\
\midrule
\midrule
GPT-4                   & 100 & 4   & 20  & 13  & 21  & 8   & 13  & 2  & 4  & 185\\
GPT-3.5                 & 28  & 2   & 36  & 4   & 5   & 11  & 14  & 2  & 3  & 105\\
Claude-3-Opus                & 11  & 0   & 32  & 5   & 23  & 16  & 24  & 4  & 4  & 119\\
Gemini-1.0-pro                 & 3   & 0   & 46  & 5   & 9   & 3   & 4   & 4  & 3  & 77\\
Ernie-3.5                 & 4   & 5   & 30  & 22  & 20  & 9   & 16  & 3  & 8  & 117\\
Qwen-turbo                   & 9   & 4   & 25  & 22  & 22  & 2   & 11  & 4  & 7  & 106\\
\midrule
Starchat-beta-16B              & 1   & 1   & 29  & 33  & 30  & 4   & 19  & 9  & 8  & 134\\
Gemma-7B                  & 1   & 29  & 33  & 15  & 17  & 1   & 10  & 0  & 2  & 108\\
Mixtral-Instruct-7B                & 1   & 26  & 11  & 42  & 21  & 9   & 19  & 6  & 11 & 146 \\
Llama2-chat-7B                  & 1   & 20  & 17  & 116 & 3   & 13  & 3   & 1  & 2  & 176\\
CodeLlama-Instruct-7B              & 1   & 0  & 18  & 22  & 26  & 3   & 10  & 3  & 4  & 87\\
WizardCoder-Python-7B            & 4   & 37  & 30  & 10  & 11  & 6   & 21  & 2  & 2  & 123\\
DeepseekCoder-Instruct-6.7B          & 8   & 2   & 37  & 10  & 8   & 3   & 8   & 0  & 2  & 78\\
Magicoder-S-DS-6.7B             & 3   & 2   & 32  & 5   & 17  & 6   & 5   & 1  & 4  & 75\\
ChatGLM3-6B                & 2   & 2  & 27  & 36  & 13  & 3   & 19  & 1  & 4  & 107\\
Codegeex2-6B              & 1   & 38  & 33  & 11  & 15  & 3   & 9   & 2  & 5  & 117\\
\midrule
\textbf{SUM}                    & 178 & 172 & 456 & 371 & 261 & 100 & 205 & 44 & 73 & 1860\\ 
\bottomrule
\end{tabular}
\caption{Statistics of solution codes in our CJ-Eval benchmark. Each row indicates the large language model that generates the solution code, and each column represents the choice of the code on the corresponding problem. "\textbf{SUM}" denotes the sum of data in the respective row or column. Our CJ-Eval benchmark contains a total of 1,860 solution codes. } 
\label{tab:statistics}
\end{table*}
\begin{table*}[t]
\centering
\begin{minipage}[t]{0.5\textwidth}
\centering
\renewcommand{\arraystretch}{1.1}
\setlength{\tabcolsep}{3pt}
\begin{tabular}{c|cccccc|c}
\Xhline{1px}
\multirow{2}{*}{Model} & \multicolumn{6}{c|}{Verdict}      & \multirow{2}{*}{SUM} \\ \cline{2-7}
                       & A   & B   & C   & D   & E   & F   &                      \\ \hline
gpt4                   & 100 & 4   & 20  & 13  & 8   & 40  & 185                  \\
gpt3.5                 & 28  & 2   & 36  & 4   & 11  & 24  & 105                  \\
claude3                & 11  & 0   & 32  & 5   & 16  & 55  & 119                  \\
gemini                 & 3   & 0   & 46  & 5   & 3   & 20  & 77                   \\
wenxin                 & 4   & 5   & 30  & 22  & 9   & 47  & 117                  \\
qwen                   & 9   & 4   & 25  & 22  & 2   & 44  & 106                  \\
chatglm                & 2   & 2   & 27  & 36  & 3   & 37  & 107                  \\
codegeex               & 1   & 38  & 33  & 11  & 3   & 31  & 117                  \\
codellama              & 1   & 0   & 18  & 22  & 3   & 43  & 87                   \\
deepseek          & 8   & 2   & 37  & 10  & 3   & 18  & 78                   \\
gemma                  & 1   & 29  & 33  & 15  & 1   & 29  & 108                  \\
llama                  & 1   & 20  & 17  & 116 & 13  & 9   & 176                  \\
magiccoder             & 3   & 2   & 32  & 5   & 6   & 27  & 75                   \\
Mixtral                & 1   & 26  & 11  & 42  & 9   & 57  & 146                  \\
starcoder              & 1   & 1   & 29  & 33  & 4   & 66  & 134                  \\
wizardcoder            & 4   & 37  & 30  & 10  & 6   & 36  & 123                  \\ \hline
SUM                    & 178 & 172 & 456 & 371 & 100 & 583 & 1860                 \\
\Xhline{1px}
\end{tabular}
\caption{Statistics of solution codes in the middle setting of CJ-Eval benchmark.}
\label{tab:middle_stat_dataset}
\end{minipage}
\hspace{1cm}
\begin{minipage}[t]{0.42\textwidth}
\centering
\renewcommand{\arraystretch}{1.1}
\setlength{\tabcolsep}{3pt}
\begin{tabular}{c|ccc|c}
\Xhline{1px}
\multirow{2}{*}{Model} & \multicolumn{3}{c|}{Verdict} & \multirow{2}{*}{SUM} \\ \cline{2-4}
                       & A       & B       & C        &                      \\ \hline
gpt4                   & 100     & 4       & 81       & 185                  \\
gpt3.5                 & 28      & 2       & 75       & 105                  \\
claude3                & 11      & 0       & 108      & 119                  \\
gemini                 & 3       & 0       & 74       & 77                   \\
wenxin                 & 4       & 5       & 108      & 117                  \\
qwen                   & 9       & 4       & 93       & 106                  \\
chatglm                & 2       & 2       & 103      & 107                  \\
codegeex               & 1       & 38      & 78       & 117                  \\
codellama              & 1       & 0       & 86       & 87                   \\
deepseek          & 8       & 2       & 68       & 78                   \\
gemma                  & 1       & 29      & 78       & 108                  \\
llama                  & 1       & 20      & 155      & 176                  \\
magiccoder             & 3       & 2       & 70       & 75                   \\
Mixtral                & 1       & 26      & 119      & 146                  \\
starcoder              & 1       & 1       & 132      & 134                  \\
wizardcoder            & 4       & 37      & 82       & 123                  \\ \hline
SUM                    & 178     & 172     & 1510     & 1860                 \\
\Xhline{1px}
\end{tabular}
\caption{Statistics of solution codes in the easy setting of CJ-Eval benchmark.}
\label{tab:easy_stat_dataset}
\end{minipage}
\end{table*}

\section{Algorithm for Filtering by Verdict}
\label{sec:algo_filter}
We introduce our filter by verdict method in Algorithm~\ref{alg:process}. The primary purpose of the filtering is to ensure that for each problem, there is at most one solution code per verdict. Additionally, we aim for an even distribution of the source models for the codes. After filtering, we initially obtained 1,994 data points. However, we later discovered that 134 of these data points, labeled as B (compilation error), had empty code fields. Therefore, we manually removed these 134 data points to obtain the final 1,860 data points mentioned in Section~\ref{sec:filter}.

\section{Statistics across Different Settings}
\label{sec:stat_easy_middle}

For the hard setting, the statistical information regarding the models and verdicts corresponding to these solution codes is provided in Table~\ref{tab:statistics}. It can be observed that the distribution of solution codes across source models and choices is relatively uniform. One exception is that the majority of solution codes with choice A come from GPT-4. This is because most of the accepted codes obtained during code generation are produced by GPT-4.

Table~\ref{tab:middle_stat_dataset} and Table~\ref{tab:easy_stat_dataset} show the statistics of our middle setting and easy setting, respectively. In the middle setting, choice F is the sum of choices \verb|EGHI| in the hard setting. In the easy setting, choice C is the sum of choices C to I in the hard setting. It can be observed that the data in the middle setting remains relatively balanced, but the easy setting faces a severe class imbalance issue. Therefore, we introduced the Macro-F1 metric, which is more robust to class imbalance, for a more comprehensive evaluation.

\section{Model Links}\label{app:links}

\begin{itemize}
    \item mistralai/Mistral-Nemo-Instruct: \url{https://huggingface.co/mistralai/Mistral-Nemo-Instruct-2407}
    \item google/gemma-2-9b-it: \url{https://huggingface.co/google/gemma-2-9b-it}
    \item meta-llama/Meta-Llama-3.1-8B-Instruct: \url{https://huggingface.co/meta-llama/Meta-Llama-3.1-8B-Instruct}
    \item Qwen/Qwen2-7B-Instruct: \url{https://huggingface.co/Qwen/Qwen2-7B-Instruct}
    \item Qwen/CodeQwen1.5-7B-Chat: \url{https://huggingface.co/Qwen/CodeQwen1.5-7B-Chat}
    \item meta-llama/CodeLlama-7b-Instruct-hf: \url{https://huggingface.co/meta-llama/CodeLlama-7b-Instruct-hf}
    \item google/codegemma-7b-it: \url{https://huggingface.co/google/codegemma-7b-it}
    \item deepseek-ai/deepseek-coder-6.7b-instruct: \url{https://huggingface.co/deepseek-ai/deepseek-coder-6.7b-instruct}
\end{itemize}

\begin{table*}[ht]
    \centering
    \begin{tabular}{lccccc}
        \toprule
        \textbf{Model} & \textbf{Size} & \textbf{Shot} & \textbf{Easy} & \textbf{Middle} & \textbf{Hard}\\
        \midrule
        \midrule
        \rowcolor{pink!50}
        \multicolumn{6}{c}{\textit{Proprietary Models}}\\
        \multirow{2}{*}{\Openaiemoji{}~\textbf{GPT-4o}} & \multirow{2}{*}{-} & 0 & 0.16 & 0.16 & 0.81 \\
        & & 1 & 0.0 & 0.0 & 0.0\\
        \midrule
        \multirow{2}{*}{\Anthropicemoji{}~\textbf{Claude-3.5-Sonnet}} & \multirow{2}{*}{-} & 0 & 0.0 & 0.0 & 0.0\\
        & & 1 & 0.0 & 0.0 & 0.0\\
        \midrule
        \multirow{2}{*}{\Googleemoji{}~\textbf{Gemini-1.5-Pro}} & \multirow{2}{*}{-} & 0 & 1.02 & 3.87 & 3.87 \\
        & & 1 & 0.0 & 0.0 & 0.0 \\
        \midrule
        \multirow{2}{*}{\Openaiemoji{}~\textbf{GPT-3.5-turbo}} & \multirow{2}{*}{-} & 0 & 0.11 & 0.16 & 0.32 \\
        & & 1 & 0.0 & 0.0 & 0.0 \\
        \midrule
        \rowcolor{green!30}
        \multicolumn{6}{c}{\textit{Open-Source Generalist Models}}\\
        \multirow{2}{*}{\Mistralemoji{}~\textbf{Mistral-Nemo-Instruct}} & \multirow{2}{*}{12B} & 0 & 2.37 & 2.47 & 1.45\\
        & & 1 & 0.32 & 0.38 & 0.27 \\
        \midrule
        \multirow{2}{*}{\Googleemoji{}~\textbf{Gemma2-IT}} & \multirow{2}{*}{9B} & 0 & 0.22 & 0.81 & 0.38 \\
        & & 1 & 0.43 & 0.27 & 0.38 \\
        \midrule
        \multirow{2}{*}{\Metaemoji{}~\textbf{Llama-3.1-Instruct}} & \multirow{2}{*}{8B} & 0 & 8.06 & 37.42 & 48.23 \\
        & & 1 & 11.34 & 8.71 & 12.15 \\
        \midrule
        \multirow{2}{*}{\Qwenemoji{}~\textbf{Qwen2-Instruct}} & \multirow{2}{*}{7B} & 0 & 10.16 & 1.99 & 0.70\\
        & & 1 & 1.34 & 0.0 & 0.0\\
        \midrule
        \rowcolor{green!30}
        \multicolumn{6}{c}{\textit{Open-Source Specific Code Models}}\\
        \multirow{2}{*}{\Qwenemoji{}~\textbf{CodeQwen1.5-Chat}} & \multirow{2}{*}{7B} & 0 & 0.0 & 0.0 & 0.0 \\
        & & 1 & 0.05 & 0.11 & 0.11 \\
        \midrule
        \multirow{2}{*}{\Metaemoji{}~\textbf{CodeLlama-Instruct}} & \multirow{2}{*}{7B} & 0 & 2.37 & 2.47 & 1.45 \\
        & & 1 & 0.32 & 0.38 & 0.27 \\
        \midrule
        \multirow{2}{*}{\Googleemoji{}~\textbf{CodeGemma-IT}} & \multirow{2}{*}{7B} & 0 & 44.41 & 58.71 & 60.22 \\
        & & 1 & 37.8 & 40.7 & 36.02 \\
        \midrule
        \multirow{2}{*}{\Deepseekemoji{}~\textbf{DeepseekCoder-Instruct}} & \multirow{2}{*}{6.7B} & 0 & 0.22 & 0.16 & 0.75 \\
        & & 1 & 2.20 & 1.24 & 8.49 \\
        \bottomrule
    \end{tabular}
    \caption{Failure rates (in percentages) of evaluated models.}
    \label{tab:fail_case}
\end{table*}

\section{Discussion on Failure Cases}
\label{sec:fail_case}

We refer to the situation where the model fails to successfully generate an answer (choice) as failing cases. The percentage of failing cases for each model is presented in Table~\ref{tab:fail_case}. It can be observed that most models have a low or zero failure rate, but a few models (Llama-3.1 and DeepseekCoder) exhibit high failure rates. Using a 1-shot prompt can help mitigate the issue of high failure rates.

Given the low failure rates of the proprietary models, it demonstrates that our prompt and answer extraction processes are well designed. The higher failure rates observed in open-source models are likely due to their weaker ability to understand the prompts. Therefore, we do not further attempt to reduce the failure rates of the open-source models.

Common failing cases include:
\begin{itemize}
    \item Model refusing to answer the question. Example from DeepseekCoder: "I'm sorry, but I can't provide the answer to this question as it's not related to computer science. \ldots".
    \item Model attempting to write code or rephrase the problem statement instead of judging. Example from Llama-3.1: "def max\_sections(n, q, intervals): \ldots".
    \item Model not following the specified format. Example from CodeGemma: "The code is not well-written and has a number of issues. The function $\ldots$".
\end{itemize}

\section{Full Results}\label{app:all_results}

\begin{table*}[htbp]
    \centering
    \begin{tabular}{lcccccccc}
        \toprule
        \multirow{2}{*}{\textbf{Model}} & \multirow{2}{*}{\textbf{Size}} & \multirow{2}{*}{\textbf{Shot}} & \multicolumn{2}{c}{\textbf{Easy}} & \multicolumn{2}{c}{\textbf{Middle}} & \multicolumn{2}{c}{\textbf{Hard}}\\
        & & &\textbf{Acc} & \textbf{F1} & \textbf{Acc} & \textbf{F1} & \textbf{Acc} & \textbf{F1}\\
        \midrule
        \midrule
        \rowcolor{gray!40}
        \multicolumn{9}{c}{\textit{Simple Strategies}}\\
        \textbf{Random} & - & - & 33.76 & 25.65 & 16.29 & \textbf{15.08} & 12.31 & \textbf{10.75} \\
        \textbf{Always AC} & - & - & 9.57 & 5.82 & 9.57 & 2.91 & 9.96 & 2.26 \\
        \textbf{Always Most Frequent Choice} & - & - & \textbf{81.18} & \textbf{29.87} & \textbf{31.34} & 7.95 & \textbf{25.52} & 5.08 \\
        \midrule
        \rowcolor{pink!50}
        \multicolumn{9}{c}{\textit{Proprietary Models}}\\
        \multirow{3}{*}{\Openaiemoji{}~\textbf{GPT-4o}} & \multirow{3}{*}{-} & 0 & 84.30 & 38.16 & 31.56 & 20.67 & 30.75 & 13.61\\
        & & 1 & \textbf{84.73} & 52.52 & 32.80 & 23.82 & 31.99 & 15.06\\
        & & 1 (CoT) & 79.89 & 39.97 & 36.45 & 29.12 & 34.30 & 21.62\\
        \midrule
        \multirow{3}{*}{\Anthropicemoji{}~\textbf{Claude-3.5-Sonnet}} & \multirow{3}{*}{-} & 0 & 80.11 & 50.83 & 31.18 & 27.02 & 30.86 & 19.05\\
        & & 1 & 81.67 & \textbf{55.68} & 33.76 & 31.73 & 32.31 & 20.50\\
        & & 1 (CoT) & 77.85 & 45.05 & \textbf{39.46} & \textbf{36.76} & \textbf{35.22} & \textbf{27.91}\\
        \midrule
        \multirow{2}{*}{\Googleemoji{}~\textbf{Gemini-1.5-Pro}} & \multirow{2}{*}{-} & 0 & 80.38 & 33.91 & 31.29 & 22.65 & 28.39 & 15.76\\
        & & 1 & 80.97 & 43.02 & 32.63 & 27.05 & 31.61 & 18.92\\
        \midrule
        \multirow{3}{*}{\Openaiemoji{}~\textbf{GPT-3.5-turbo}} & \multirow{3}{*}{-} & 0 & 38.06 & 18.68 & 16.24 & 10.31 & 12.63 & 5.83\\
        & & 1 & 62.04 & 35.62 & 29.95 & 16.78 & 12.69 & 7.94\\
        & & 1 (CoT)& 66.61 & 39.21 & 24.95 & 17.72 & 17.90 & 12.52\\
        \midrule
        \rowcolor{green!30}
        \multicolumn{9}{c}{\textit{Open-Source Generalist Models}}\\
        \multirow{2}{*}{\Mistralemoji{}~\textbf{Mistral-Nemo-Instruct}} & \multirow{2}{*}{12B} & 0 & 9.62 & 4.55 & 9.46 & 2.52 & 9.52 & 1.76\\
        & & 1 & 30.75 & 15.44 & 10.54 & 4.47 & 10.27 & 2.98\\
        \midrule
        \multirow{2}{*}{\Googleemoji{}~\textbf{Gemma2-IT}} & \multirow{2}{*}{9B} & 0 & 57.04 & 19.80 & 19.14 & 9.30 & \textbf{18.87} & \textbf{9.17}\\
        & & 1 & 51.83 & 23.04 & 15.86 & 8.50 & 16.08 & 6.94\\
        \midrule
        \multirow{2}{*}{\Metaemoji{}~\textbf{Llama-3.1-Instruct}} & \multirow{2}{*}{8B} & 0 & 13.01 & 11.81 & 10.11 & 9.74 & 9.03 & 7.69\\
        & & 1 & 10.05 & 10.20 & 10.48 & 6.73 & 8.76 & 4.19\\
        \midrule
        \multirow{2}{*}{\Qwenemoji{}~\textbf{Qwen2-Instruct}} & \multirow{2}{*}{7B} & 0 & 21.88 & 14.51 & 16.99 & 7.56 & 9.89 & 3.51\\
        & & 1 & \textbf{66.40} & \textbf{30.29} & \textbf{28.28} & \textbf{11.39} & 8.82 & 5.31\\
        \midrule
        \rowcolor{green!30}
        \multicolumn{9}{c}{\textit{Open-Source Code Models}}\\
        \multirow{2}{*}{\Qwenemoji{}~\textbf{CodeQwen1.5-Chat}} & \multirow{2}{*}{7B} & 0 & 15.05 & 13.03 & 9.89 & 3.95 & \textbf{10.00} & \textbf{3.37}\\
        & & 1 & 11.67 & 11.16 & \textbf{10.11} & 4.04 & \textbf{10.00} & 2.61\\
        \midrule
        \multirow{2}{*}{\Metaemoji{}~\textbf{CodeLlama-Instruct}} & \multirow{2}{*}{7B} & 0 & \textbf{59.84} & \textbf{21.15} & 5.16 & 3.78 & 5.48 & 3.13\\
        & & 1 & 7.85 & 4.56 & 8.01 & 2.49 & 7.58 & 1.68\\
        \midrule
        \multirow{2}{*}{\Googleemoji{}~\textbf{CodeGemma-IT}} & \multirow{2}{*}{7B} & 0 & 16.40 & 10.39 & 5.48 & 3.69 & 5.59 & 3.17\\
        & & 1 & 9.78 & 6.76 & 6.77 & 3.58 & 7.58 & 2.43\\
        \midrule
        \multirow{2}{*}{\Deepseekemoji{}~\textbf{DeepseekCoder-Instruct}} & \multirow{2}{*}{6.7B} & 0 & 10.38 & 7.28 & 9.73 & 2.80 & 9.68 & 1.97\\
        & & 1 & 15.05 & 8.59 & 10.05 & \textbf{4.10} & 9.78 & 2.48\\
        \bottomrule
    \end{tabular}
    \caption{More results on our CJ-Eval benchmark. We present all results for the 0-shot, 1-shot, and 1-shot Chain-of-Thought settings.}
    \label{tab:all_results}
\end{table*}

Due to space limit, we do not included the full 1-shot and 1-shot CoT results in the main text. The complete results can be found in Table~\ref{tab:all_results}. Among all methods, Claude-3.5 achieve the best performance in both the middle and hard settings when using the 1-shot CoT example, significantly outperforming other methods. For the easy setting, GPT-4o and Claude-3.5 achieve the best performance in terms of accuracy and Macro F1, respectively, under the 1-shot example.

\section{Prompt}
\label{sec:prompt}

We demonstrate our prompts in this section.

\paragraph{General Prompt Template.}

We will introduce zero-shot, one-shot, one-shot CoT, and few-shot prompts in this section. Before that, our prompts utilize the same prompt template, as shown in Figure~\ref{fig:prompt1}. The four types of prompts differ only in the content filled in the "Placeholder."

Placeholder A is used to insert the description of the choices. For different difficulties settings in evaluation (i.e., easy, middle, and hard), we design different choices description to be filled into the first curly bracket (Placeholder A) in the template. The choices for the three settings, from easy to hard, are as follows.

For Placeholder B, in the one-shot, one-shot CoT, and few-shot prompts, it is filled with one or more examples correspondingly. In the zero-shot prompt, this content is empty.

Placeholders C and D are the same for all four types of prompts, containing the description of the new problem and the corresponding solution code to be judged.

\lstset{
    backgroundcolor=\color[RGB]{245,245,244},
    breaklines=true,
    breakindent=0px,
    basicstyle=\ttfamily\small
}\begin{lstlisting}
(A). AC
(B). CE
(C). Not AC

(A). AC
(B). CE
(C). Not AC, only WA errors
(D). Not AC, only RE errors
(E). Not AC, only TLE errors
(F). Not AC for at least two types of errors

(A). AC
(B). CE
(C). Not AC, only WA errors
(D). Not AC, only RE errors
(E). Not AC, both WA and RE errors
(F). Not AC, only TLE errors
(G). Not AC, both WA and TLE errors
(H). Not AC, both TLE and RE errors
(I). Not AC, all WA, RE, and TLE errors
\end{lstlisting}

\paragraph{Zero-shot Prompt.}

As described above, a complete zero-shot prompt, with Figure~\ref{fig:prompt1} as the template, involves inserting the corresponding choices description based on difficulty at Placeholder A, leaving Placeholder B empty, and filling in the questions to be answered at Placeholders C and D.

\paragraph{One-shot Prompt.}

The only difference between a one-shot prompt and a zero-shot prompt is that a single example question is inserted at Placeholder B in the one-shot prompt. This same one-shot example question is used consistently throughout the evaluation. The full example is shown in Figure~\ref{fig:prompt2}.

Note that the answer to the example question depends on the difficulty setting of the current evaluation, as the meanings of the options vary with difficulty. The answer to this question is A for easy, middle, and hard difficulty levels.

\begin{figure*}[h]
\lstset{
    backgroundcolor=\color[RGB]{245,245,244},
    breaklines=true,
    breakindent=0px,
    basicstyle=\ttfamily\small
}\begin{lstlisting}
# Task Requirement

You need to check whether the following code can pass the given programming problem, which may come from interview questions or competitive programming problems on sites like LeetCode, Codewars, Codeforces, or Kattis. You need to comprehensively consider various test cases, assuming that the test cases are sufficient to detect any existing errors.

## Explanation of Choices

The meanings of some results are explained as follows:
- AC: Accepted, the program is completely correct and passes all hidden test cases;
- CE: Compilation Error, detected by Python before the program runs (e.g., mismatched parentheses, incorrect - indentation, invalid Python syntax, etc.);
- WA: Wrong Answer, the output is incorrect for at least one test case;
- RE: Runtime Error, the program crashes for at least one test case;
- TLE: Time Limit Exceeded, the program exceeds the time limit (2 seconds per test case) for at least one test case, within the problem's constraints.

## Choices

Please select the option that correctly describes the result (select only one option). You must directly output the answer by a single letter.

{Placeholder A: The choices and their explanation are provided here. The choices vary depending on whether the current evaluation setting is Easy, Middle, or Hard.}

{Placeholder B: When the evaluation setting is 1-shot, an example problem with a solution code will be provided here. Otherwise, nothing will be inserted.}

## New Problem

### New Problem Description

{Placeholder C: The textual description of the problem to be judged and a few input-output examples are provided here.}

### New Solution to be Judged

{Placeholder D: The Python solution code to be judged is provided here.}

### New Answer
You must directly output the answer by a single letter.
The answer is 
\end{lstlisting}
\caption{The template of our prompt. The content enclosed in "\{\}" (marked as ``Placeholder'') will be filled in later based on the evaluation setting and the problem with the judged solution code.}
\label{fig:prompt1}
\end{figure*}

\begin{figure*}[h]
\lstset{
    backgroundcolor=\color[RGB]{245,245,244},
    breaklines=true,
    breakindent=0px,
    basicstyle=\ttfamily\small
}\begin{lstlisting}
## Example problem

### Example Problem Description

Finally, the pandemic is over in ChefLand, and the chef is visiting the school again. Chef likes to climb the stairs of his school's floor by skipping one step, sometimes chef climbs the stairs one by one. Simply, the chef can take one or 2 steps in one upward movement. There are N stairs between ground and next floor. The chef is on the ground floor and he wants to go to the next floor with Cheffina but, Cheffina asks chef in how many ways, the chef can reach the next floor normally or any combination of skipping one step, where order doesn't matter. 

-----Input:-----

- First-line will contain $T$, the number of test cases. Then the test cases follow. 
- Each test case contains a single line of input, two integers $N$. 

-----Output:-----

For each test case, output in a single line answer as the number of ways.

-----Constraints-----
- $1 \leq T \leq 1000$
- $1 \leq N \leq 10^5$

-----Sample Input:-----
1
3

-----Sample Output:-----
2

-----EXPLANATION:-----
ways: [1,1,1], here chef climb to the next floor, one by one stair.
[1,2], here chef climb to the next floor, one step first and after that 2 stairs at once.
Note, [2,1] consider the same as that of [1,2] hence ignored.

### Example Solution to be Judged

def count_ways(n):
    return (n // 2) + 1

def solve(test_cases):
    results = []
    for n in test_cases:
        results.append(count_ways(n))
    return results

import sys
input = sys.stdin.read
data = input().split()

T = int(data[0])
test_cases = [int(data[i]) for i in range(1, T + 1)]

results = solve(test_cases)

for result in results:
    print(result)

### Example Answer
You must directly output the answer by a single letter.
The answer is (AAA).
\end{lstlisting}
\caption{The example problem to fill in the prompt template in the one-shot setting. The ``AAA'' in the last parentheses represents the content inserted under hard, medium, and easy settings, respectively.}
\label{fig:prompt2}
\end{figure*}

\paragraph{One-shot CoT Prompt.}
Similar to a one-shot prompt, the primary difference with a one-shot CoT (Chain-of-Thought) prompt lies in the example provided, which includes a detailed Chain-of-Thought process. The complete example is shown in Figure~\ref{fig:cot_prompt}, with the answers to the sample question being C, E, and F under the easy, middle, and hard settings, respectively. The prompt also includes slight modifications to guide the model to think step by step. These specific alterations can be found in our released CJ-Eval Benchmark.

\begin{figure*}[htbp]
\lstset{
    backgroundcolor=\color[RGB]{245,245,244},
    breaklines=true,
    breakindent=0px,
    basicstyle=\ttfamily\scriptsize
}\begin{lstlisting}
## Example problem

### Example Problem Description

You have an array $a_1, a_2, \dots, a_n$. 
Let's call some subarray $a_l, a_{l + 1}, \dots , a_r$ of this array a subpermutation if it contains all integers from $1$ to $r-l+1$ exactly once. For example, array $a = [2, 2, 1, 3, 2, 3, 1]$ contains $6$ subarrays which are subpermutations: $[a_2 \dots a_3]$, $[a_2 \dots a_4]$, $[a_3 \dots a_3]$, $[a_3 \dots a_5]$, $[a_5 \dots a_7]$, $[a_7 \dots a_7]$.
You are asked to calculate the number of subpermutations.


-----Input-----

The first line contains one integer $n$ ($1 \le n \le 3 \cdot 10^5$).
The second line contains $n$ integers $a_1, a_2, \dots , a_n$ ($1 \le a_i \le n$). 
This array can contain the same integers.


-----Output-----

Print the number of subpermutations of the array $a$.

-----Examples-----
Input
8
2 4 1 3 4 2 1 2
Output
7

Input
5
1 1 2 1 2
Output
6


-----Note-----

There are $7$ subpermutations in the first test case. Their segments of indices are $[1, 4]$, $[3, 3]$, $[3, 6]$, $[4, 7]$, $[6, 7]$, $[7, 7]$ and $[7, 8]$.
In the second test case $6$ subpermutations exist: $[1, 1]$, $[2, 2]$, $[2, 3]$, $[3, 4]$, $[4, 4]$ and $[4, 5]$.

### Example Solution to be Judged

def count_subpermutations(n, arr):
    n = len(arr)
    count = 0
    for i in range(n):
        nxt = 1
        st = set()
        for j in range(i,-1,-1):
            st.add(arr[j])
            while (nxt in st):
                nxt += 1
            if nxt==i-j+1+1:
                count += 1
    return count

n = int(input())
arr = list(map(int, input().split()))
print(count_subpermutations(n,arr))

### Example Answer
You must analyze the code first and then directly output the answer following the given format at the end of your response.

This code enumerates the right endpoint of the interval as i and attempts to verify whether j, as the left endpoint of the interval, is valid. The code stores the numbers within the interval [i, j] in a set and then finds the smallest missing number. If this number equals the interval length plus one, it indicates that all numbers from 1 to the interval length appear exactly once, making the interval valid. The logic of the code is correct and will not result in a WA. The implementation is also error-free, preventing RE. However, the algorithm's complexity is O(n^2), which cannot handle the given data range and will result in a TLE.

The answer is (FEC).
\end{lstlisting}
\caption{The example problem to fill in the prompt template in the one-shot CoT setting. The ``FEC'' in the last parentheses represents the content inserted under hard, medium, and easy settings, respectively.}
\label{fig:cot_prompt}
\end{figure*}

\paragraph{Few-shot Prompt.}

Few-shot prompts refer to the 2-shot and 3-shot settings depicted in Figure~\ref{fig:ana_few_shot}. The main difference here is that we include 2 or 3 examples at Placeholder B. These newly introduced examples can also be found in our released CJ-Eval Benchmark.

\section{Analysis for Chain-of-Thought}
\label{sec:cot_analysis}

To investigate whether the improvements brought by Chain-of-Thought (CoT) stem from a clearer reasoning of the problem, we present the output of Claude-3.5-Sonnet on a code judging task with a one-shot CoT example. The problem is shown in Figure~\ref{fig:cot_input}, and the model's response is displayed in Figure~\ref{fig:cot_output}. The results indicate that the model identified several errors in the code and discussed whether the code would result in WA (Wrong Answer), CE (Compilation Error), RE (Runtime Error), or TLE (Time Limit Exceeded) errors, ultimately providing the correct answer.

\begin{figure*}[htbp]
\lstset{
    backgroundcolor=\color[RGB]{245,245,244},
    breaklines=true,
    breakindent=0px,
    basicstyle=\ttfamily\scriptsize
}\begin{lstlisting}
### New Problem Description

An accordion is a string (yes, in the real world accordions are musical instruments, but let's forget about it for a while) which can be represented as a concatenation of: an opening bracket (ASCII code $091$), a colon (ASCII code $058$), some (possibly zero) vertical line characters (ASCII code $124$), another colon, and a closing bracket (ASCII code $093$). The length of the accordion is the number of characters in it.

For example, [::], [:||:] and [:|||:] are accordions having length $4$, $6$ and $7$. (:|:), {:||:}, [:], ]:||:[ are not accordions. 

You are given a string $s$. You want to transform it into an accordion by removing some (possibly zero) characters from it. Note that you may not insert new characters or reorder existing ones. Is it possible to obtain an accordion by removing characters from $s$, and if so, what is the maximum possible length of the result?


-----Input-----
The only line contains one string $s$ ($1 \le |s| \le 500000$). It consists of lowercase Latin letters and characters [, ], : and |.

-----Output-----
If it is not possible to obtain an accordion by removing some characters from $s$, print $-1$. Otherwise print maximum possible length of the resulting accordion.

-----Examples-----
Input
|[a:b:|]
Output
4

Input
|]:[|:]
Output
-1

### New Solution to be Judged

def max_accordion_length(s):
    # Initialize the stack and the max length
    stack = []
    max_len = 0

    # Iterate over each character in the string
    for c in s:
        # If the character is an opening bracket or a colon, push it to the stack
        if c == '[' or c == ':':
            stack.append(c)
        # If the character is a vertical line, check if the stack is not empty and the top element is a colon
        elif c == '|' and stack and stack[-1] == ':':
            stack.append(c)
        # If the character is a closing bracket, pop elements from the stack until an opening bracket is found
        elif c == ']' and stack and stack[-1] == '[':
            stack.pop()
            # If the stack is not empty, it means we have a complete accordion
            if stack:
                max_len = max(max_len, len(stack))
        # If the character is a colon, check if the stack is not empty and the top element is a colon
        elif c == ':' and stack and stack[-1] == ':':
            stack.append(c)
        # If the character is not a valid accordion character, return -1
        else:
            return -1

    # If the stack is not empty, it means we have an incomplete accordion
    if stack:
        return -1

    # Return the max length of the accordion
    return max_len

# Test the function
print(max_accordion_length(input()))


### New Answer
You must analyze the code first and then directly output the answer following the given format at the end of your response.
\end{lstlisting}
\caption{A solution code and its corresponding problem to be judged in the CoT setting.}
\label{fig:cot_input}
\end{figure*}

\begin{figure*}[htbp]
\lstset{
    backgroundcolor=\color[RGB]{245,245,244},
    breaklines=true,
    breakindent=0px,
    basicstyle=\ttfamily\small
}\begin{lstlisting}
1. The function `max_accordion_length` takes a string as input and uses a stack to keep track of the accordion structure.

2. It iterates through each character in the string:
   - For '[' or ':', it pushes the character onto the stack.
   - For '|', it checks if the top of the stack is ':' before pushing.
   - For ']', it checks if the top of the stack is '[', then pops it and updates the max length if a complete accordion is found.
   - For ':', it checks if the top of the stack is also ':' before pushing.
   - For any other character, it returns -1.

3. If the stack is not empty at the end, it returns -1 (incomplete accordion).

4. The function returns the maximum length of a valid accordion found.

However, there are several issues with this implementation:

1. It doesn't handle the case where '[' appears after ':' correctly.
2. It doesn't properly track the start of an accordion.
3. It doesn't correctly handle nested accordions or multiple accordions in the string.
4. It may return incorrect results for some valid inputs.

For example:
- For input "[:||:]", it will return 0 instead of 6.
- For input "[::][::]", it will return 2 instead of 4.
- For input "[:||:][::]", it will return 5 instead of 6.

These issues will lead to Wrong Answer (WA) for many test cases. The code doesn't have any syntax errors that would cause Compilation Error (CE), and it's unlikely to cause Runtime Error (RE) or Time Limit Exceeded (TLE) given the problem constraints.

Therefore, the correct answer is:

(C)
\end{lstlisting}
\caption{Output from Claude-3.5-Sonnet for the task in Figure~\ref{fig:cot_input}. The answer C is correct.}
\label{fig:cot_output}
\end{figure*}

\end{document}